\documentclass[sigconf,screen]{acmart}

\usepackage[T1]{fontenc}
\usepackage{CJK}

\usepackage{makecell}
\usepackage{subfigure}

\AtBeginDocument{%
  \providecommand\BibTeX{{%
    \normalfont B\kern-0.5em{\scshape i\kern-0.25em b}\kern-0.8em\TeX}}}

\makeatletter
\def\@ACM@copyright@check@cc{}
\makeatother
\copyrightyear{2025}
\acmYear{2025}
\setcopyright{cc}
\setcctype{by-nc-sa}
\acmConference[CHI '25]{CHI Conference on Human Factors in Computing Systems}{April 26-May 1, 2025}{Yokohama, Japan}
\acmBooktitle{CHI Conference on Human Factors in Computing Systems (CHI '25), April 26-May 1, 2025, Yokohama, Japan}
\acmDOI{10.1145/3706598.3714107}
\acmISBN{979-8-4007-1394-1/25/04}

\begin{document}

\begin{CJK}{UTF8}{mj}

\title[VTuber's Atelier: The Design Space, Challenges, and Opportunities for VTubing]{VTuber's Atelier: The Design Space, Challenges, and Opportunities for VTubing}

\author{Daye Kim}
\email{kimdyae@soongsil.ac.kr}
\orcid{0009-0000-2296-6920}
\authornote{Both authors contributed equally to this work}
\affiliation{%
  \institution{The Global School of Media\\Soongsil University}
  \city{Seoul}
  \country{Republic of Korea}
}

\author{Sebin Lee}
\email{leesebin@soongsil.ac.kr}
\orcid{0000-0002-8927-0319}
\authornotemark[1]
\affiliation{%
  \institution{Department of Culture Contents\\Soongsil University}
  \city{Seoul}
  \country{Republic of Korea}
}

\author{Yoonseo Jun}
\email{iamyoonseo@soongsil.ac.kr}
\orcid{0009-0006-7209-6074}
\affiliation{%
  \institution{The Global School of Media\\Soongsil University}
  \city{Seoul}
  \country{Republic of Korea}
}

\author{Yujin Shin}
\email{tankgirl2003@soongsil.ac.kr}
\orcid{0009-0005-1148-7672}
\affiliation{%
  \institution{The Global School of Media\\Soongsil University}
  \city{Seoul}
  \country{Republic of Korea}
}

\author{Jungjin Lee}
\email{jungjinlee@ssu.ac.kr}
\orcid{0000-0003-3471-4848}
\authornote{Corresponding author}
\affiliation{%
  \institution{The Global School of Media\\Soongsil University}
  \city{Seoul}
  \country{Republic of Korea}
}

\renewcommand{\shortauthors}{Daye Kim, Sebin Lee, Yoonseo Jun, Yujin Shin \& Jungjin Lee}

\begin{abstract}
VTubing, the practice of live streaming using virtual avatars, has gained worldwide popularity among streamers seeking to maintain anonymity.
While previous research has primarily focused on the social and cultural aspects of VTubing, there is a noticeable lack of studies examining the practical challenges VTubers face in creating and operating their avatars. 
To address this gap, we surveyed VTubers’ equipment and expanded the live-streaming design space by introducing six new dimensions related to avatar creation and control. 
Additionally, we conducted interviews with 16 professional VTubers to comprehensively explore their practices, strategies, and challenges throughout the VTubing process.
Our findings reveal that VTubers face significant burdens compared to real-person streamers due to fragmented tools and the multi-tasking nature of VTubing, leading to unique workarounds.
Finally, we summarize these challenges and propose design opportunities to improve the effectiveness and efficiency of VTubing.
\end{abstract}




\maketitle

\section{Introduction}
Advances in motion capture, computer graphics (CG), and virtual reality (VR) technologies have contributed to the growing popularity of online entertainers who use virtual avatars for live streaming without revealing their true identities. 
Known as virtual streamers, virtual YouTubers, or VTubers for short, they have gained prominence, rivaling real-person streamers in terms of engagement and popularity. 
For example, VTubers dominate the top of the list of YouTubers worldwide who have received the most Super Chats, which are monetary gifts from viewers \cite{Playboard-Superchat}.
In addition, many talented VTubers amuse their fandom by releasing albums and holding concerts through record labels \cite{Lee-2023-Ju.Taime}.
From an economic perspective, the VTuber market continues to expand rapidly, valued at USD 4.4 billion in 2022 and projected to reach USD 27.6 billion by 2029 \cite{Newswire-VTuberMarket}. 
This growth underscores the importance of understanding VTubing not just as a cultural phenomenon, but also as a domain of human-computer interaction (HCI) that challenges existing paradigms of live streaming, virtual identity, and avatar interaction.

VTubers possess unique characteristics that differentiate them from real-person streamers. 
The actors behind the virtual avatars, also known as Nakanohitos, often role-play carefully designed personas that differ from their own personalities and identities \cite{Lu-2021-MoreKawaii, Bredikhina-2020-Era, Bredikhina-2022-Becoming, Wan-2024-Investigating, Turner-2022-VirtualBeing, ferreira-2022-concept}.
Some even adopt avatars of different genders, or multiple actors may portray the same avatar \cite{Lu-2021-MoreKawaii, Bredikhina-2022-Becoming, Bredikhina-2020-Era}. 
Viewers' experiences and attitudes towards VTubers are also interesting.
They tend to be more distant and more tolerant of VTubers than real-person streamers \cite{Lu-2021-MoreKawaii, Xu-2023-Understanding}. 
While previous studies have explored the social and cultural dimensions of VTubing, limited attention has been given to the practical challenges that VTubers face in creating and operating their avatars.

From an HCI perspective, VTubing represents a unique intersection of technical complexity and user interaction.
VTubers face distinct challenges, such as integrating multiple devices (e.g., motion capture and HMD devices) for avatar control, managing real-time viewer interactions while maintaining anonymity, and crafting consistent virtual identities.
Moreover, successful VTubing requires expertise in CG, such as modeling, rigging, animating, and rendering \cite{Rohrbacher-2024-VTubing, ferreira-2022-concept}.
While virtual influencers on social media platforms primarily rely on pre-rendered content for marketing purposes \cite{Conti-2022-VIs, Choudhry-2022-Felt}, VTubers interact live, introducing a layer of unpredictability and heightened technical demands.
This live format requires VTubers to simultaneously manage avatar control, audience engagement, and streaming performance, imposing substantial cognitive and physical burdens due to the continuous multi-tasking involved \cite{Tang-2021-AlterEcho}.
These technical and interactional differences necessitate a deeper exploration of how VTubers adapt to and negotiate these constraints.
By investigating VTubing workflows and the associated tools and practices, this research seeks to contribute to both HCI and virtual avatar research by providing insights into the challenges and opportunities in this emerging field.

To achieve this, the study is guided by the following research questions:
\begin{itemize}
    \item RQ1: What are the key dimensions of the design space for the current state of VTubing equipment?
    \item RQ2: What are the current practices, strategies, and challenges throughout the VTubing process?
\end{itemize}

To explore these questions, we began with desk research to identify the equipment commonly used by VTubers. 
Building on this foundation, we surveyed 18 professional VTubers and conducted follow-up interviews with 16 of them, representing diverse backgrounds, to gain deeper insights into how they prepare for streams, configure and operate their equipment, and navigate the associated challenges. 
Our findings revealed six dimensions related to avatar creation and control, which can be incorporated into the existing design space \cite{Drosos-2022-DesignSpace} of live-streaming equipment (Section \ref{sec:designspace}).
Our in-depth interviews with practitioners also comprehensively captured the diverse struggles of VTubers in terms of equipment throughout a VTubing process, including balancing creative vs. comfortable personas, managing high-quality outputs against cost efficiency, and maximizing expressiveness while minimizing cognitive and physical burdens (Sections \ref{sec:interview} and \ref{sec:discussion}).
Drawing from these, we propose design opportunities for future systems to enhance creative content production and improve the overall VTubing experience (Section \ref{sec:design opportunities}).
Beyond VTubing, our findings have broader implications for improving avatar-based communication and virtual identity management in fields such as education, remote collaboration, and online social interaction.

Our key contributions include:
\begin{enumerate}
    \item A design space for VTubing equipment based on a comprehensive analysis involving desk research and a real-world user study
    \item A user study that sheds light on VTubers’ practices, experiences, challenges, and needs regarding their equipment
    \item A discussion of design opportunities that can make VTubing more effective and efficient
\end{enumerate}
\section{Related Work}
\subsection{Live Streaming in HCI}
Live streaming has quickly become popular, with HCI research exploring the experiences of streamers and viewers \cite{Li-2019-CoPerformance, Pellicone-2017-Performing, Wu-2022-Concerned, Lu-2018-YouWatch, Hamilton-2014-StreamingOnTwitch, Wohn-2020-AudienceManagement, Wohn-2018-ExplainingViewer}. 
Pellicone and Ahn introduced the concept of live streaming as `performance play,' focusing on streamer practices to deliver successful performances \cite{Pellicone-2017-Performing}.
Li et al. examined streamer-viewer interactions as co-performances and analyzed strategies for enhancing engagement \cite{Li-2019-CoPerformance}. 
Engaging performance can attract more viewers and provide benefits such as monetary rewards \cite{Woodcock-2019-AffectiveLabor}, personal brand development \cite{Tang-2016-Meerkat}, reputation building \cite{Fietkiewicz-2018-Dreaming}, and community creation \cite{Hu-2017-WhyDoAudiences, Brundl-2016-WhyDoBroadcast}.
To sustain engaging performances, streamers employ various strategies. 
They may adopt a naturalistic attitude to emphasize authenticity \cite{Wu-2022-Concerned, Lu-2018-YouWatch, Tang-2022-DareToDream} or use aggressive commentary to create intimacy with viewers \cite{Wu-2023-Interactions}, or manage their appearance and attire to enhance appealing \cite{Freeman-2020-StreamingIdentity, Li-2020-Spontaneous, Wu-2022-Concerned}.
Streamers also use polls \cite{Hamilton-2014-StreamingOnTwitch, Li-2019-CoPerformance, Lu-2018-YouWatch} and social media interactions \cite{Jia-2020-HowtoAttract, Hamilton-2014-StreamingOnTwitch, Pellicone-2017-Performing} to boost viewer engagement during and after streams.

In addition to crafting attractive personas, streamers’ technical skills in managing equipment are crucial for delivering high-quality performances. 
Streamers use various hardware, software, and design tools to enhance the professionalism and richness of their content \cite{Pellicone-2017-Performing}.
The fidelity of these setups often depends on the streamer's technical ability \cite{Drosos-2022-DesignSpace, Freeman-2020-StreamingIdentity}. 
Previous research on streamers’ technical experiences has explored their practices and challenges and proposed design insights.
For instance, Cai and Wohn focused on real-time moderation tools, while Mallari et al. examined the use of analytics tools on platforms like Twitch and Mixer \cite{cai-2019-categorizing, Mallari-2021-Understanding}. 
Drosos and Guo provided a comprehensive review and design space of hardware and software setups for live streaming, classifying equipment into broadcasting, video, and audio dimensions according to their fidelity levels \cite{Drosos-2022-DesignSpace}. 

Independent streamers typically play multiple roles and take on different challenges on their own, including setting up and managing streams, hosting live shows, interacting with and moderating viewers.
Accordingly, HCI researchers have actively proposed various systems to improve streamers' performance and reduce their workload \cite{Fraser-2020-TemporalSegmentation, Chen-2019-Integrating, Lessel-2017-Expanding, Chung-2021-Beyond, Hammad-2023-View, Lu-2021-Streamsketch, Nebeling-2021-XRstudio, Lu-2018-Streamwiki}.
Kobs et al. introduced fine-tuned sentiment analysis for active stream text chats, demonstrating its potential to help streamers improve their performance by providing insights into audience reactions \cite{Kobs-2020-Emote}.
StoryChat, a narrative-based viewer engagement tool, was designed to enhance audience participation, promote prosocial behavior, and evaluate the system's moderation effects \cite{Yen-2023-StoryChat}.
Fraser et al. developed algorithms to streamline post-broadcast content management by automatically segmenting live-streaming videos into meaningful sessions \cite{Fraser-2020-TemporalSegmentation}.
Additionally, several studies have explored new communication channels and interaction options to expand beyond limited text-based interactions between viewers and streamers. 
These include utilizing various modalities, such as images and videos \cite{Chen-2019-Integrating, Yang-2020-Snapstream, Chung-2021-Beyond, Lu-2021-Streamsketch}, or designing custom layouts that overlay in-game information \cite{Hammad-2023-View, Lessel-2017-Expanding}.

These diverse system proposals have been enabled by extensive prior research into the practical experiences, challenges, and strategies of streamers and their audiences.
However, most of these studies have focused primarily on real-person streamers, leaving the practical experiences of VTubers underexplored.
The emerging domain of VTubers, who use avatars for self-representation, introduces additional technical skills and equipment specific to avatar creation and control \cite{ferreira-2022-concept, jiang-2023-better}. 
These unique technical demands present distinct challenges that differ from those faced by real-person streamers. 
To address this research gap, we aim to identify the equipment used by VTubers and explore their experiences in managing various tools and technologies.
Our study has the potential to inspire a variety of future system research tailored to the context of VTubing.

\subsection{Virtual Influencers and VTubers in HCI}
With the rapid advancement of CG technology, the use of virtual avatars has expanded into diverse fields, including advertising and marketing \cite{Allal-2024-Intelligent, Kim-2024-VIMarketing, Mouritzen-2024-VIMarketing, Choudhry-2022-Felt}, entertainment \cite{Lee-2023-Ju.Taime, Lee-2023-VirtualAvatarConcert,Lee-2024-Wishbowl,Lee-2022-Understanding,Kang-2021-CodeMiko,Lee-2023-Simulcast, Lu-2021-MoreKawaii}, and education \cite{Shang-2024-incidental, Susanti-2022-VTuberOL, Mohammad-2023-DuolingoHololive}.
Advanced CGI (computer-generated imagery) techniques have introduced hyper-realistic virtual influencers that are nearly indistinguishable from real people, garnering significant attention for their comparable influence \cite{Conti-2022-VIs, Choudhry-2022-Felt}.
These virtual influencers engage with the world from a first-person perspective by uploading high-quality, non-real-time images and videos on social media platforms such as Instagram and YouTube \cite{Choudhry-2022-Felt}.
The growing use of virtual influencers in advertising and marketing has highlighted public perceptions of virtual versus real influencers, becoming a prominent topic in HCI research \cite{Allal-2024-Intelligent, Conti-2022-VIs, Lee-2024-VIvsHI, Belanche-2024-Human}.
Meanwhile, advancements in real-time rendering technology and affordable motion capture systems have led to the emergence of a unique group of users: VTubers, who utilize avatars for live streaming, setting them apart from virtual influencers \cite{BBC-virtualVloggers, What'stheDifference}.

Recently, VTubing has become one of the most popular forms of live streaming \cite{Lu-2021-MoreKawaii, Turner-2022-VirtualBeing}.
To understand this emerging phenomenon, HCI researchers have studied the differences in viewer perceptions between real-person streamers and VTubers \cite{Sakuma-2023-YouTubervVTubers, Hsieh-2019-Effectiveness}, as well as the factors that shape the VTubing viewing experience \cite{Lu-2021-MoreKawaii, Xu-2023-Understanding, Lee-2023-Ju.Taime, Miranda-2024-VTubers, Tan-2023-MoreAttached}.
For example, Lu et al. provided a detailed analysis of the factors that engage VTuber viewers, highlighting differences from real-person streamers and perceptions of Nakanohito \cite{Lu-2021-MoreKawaii}. 
Xu and Niu explored psychological factors influencing VTubing viewers, such as the perceived attractiveness of virtual identities, immersion, and psychological distance \cite{Xu-2023-Understanding}, and Lee and Lee focused on the factors affecting VTuber fandom experiences during virtual concerts \cite{Lee-2023-Ju.Taime}.

In addition to the viewer perspectives, researchers have examined VTubers' experiences in constructing and performing virtual identities \cite{Wan-2024-Investigating, Bredikhina-2020-Era, Bredikhina-2022-Becoming, Rohrbacher-2024-VTubing, Chen-2024-Host}.
For example, Wan and Lu explored how Chinese VTubers construct their virtual personas and express gender through avatars \cite{Wan-2024-Investigating}.
Bredikhina and Giard studied the babiniku phenomenon, where male VTubers use female avatars, and its effects on male self-perception \cite{Bredikhina-2022-Becoming}.
Similarly, Bredikhina, along with Rohrbacher and Mishra, explored how VTubers construct their virtual identities across different cultural contexts \cite{Bredikhina-2020-Era, Rohrbacher-2024-VTubing}.
These studies provide valuable insights into identity construction and self-representation, closely tied to the experiences of avatar users in virtual worlds \cite{Pace-2009-Socially, Neustaedter-2009-Presenting, Ducheneaut-2009-Body, McArthur-2015-AvatarAffordance}.
However, there is limited understanding of the equipment VTubers use to express themselves and their experiences and challenges they face in operating equipment.
To better understand and support VTubers’ self-expression through avatars, it is essential to address the equipment they use and their experiences with it.

\subsection{VTubing Technologies}
VTubers, like VR and CG creators, produce their content through workflows involving modeling, rigging, motion capture, animating, and rendering, often using overlapping software and hardware tools \cite{ferreira-2022-concept, jiang-2023-better}.
Similar to our study, HCI researchers have examined current AR/VR/CG tools and their users to better understand the practical experiences, challenges, and practices associated with these creative production processes, while also proposing novel design opportunities \cite{Nebeling-2018-Trouble, Ashtari-2020-Creating, Krauss-2021-Current, Cheon-2024-Cretive}.
Nebeling and Speicher reviewed and categorized existing AR/VR authoring tools based on their fidelity levels and the user skills required, offering a comprehensive overview of the AR/VR tool landscape \cite{Nebeling-2018-Trouble}.
Building on this foundation, Ashtari et al. and Krau{\ss} et al. identified that the fragmented nature of authoring tools not only complicates the process of understanding and selecting tools within the AR/VR landscape but also increases the complexity of collaborative workflows among developers \cite{Ashtari-2020-Creating, Krauss-2021-Current}.
Furthermore, Cheon and Xu conducted interviews with 14 professional motion capture actors, uncovering how the limitations of current motion capture equipment and technologies exacerbate physical constraints and challenges in their workflows \cite{Cheon-2024-Cretive}.

Although VTubers share similar workflows, technologies, and equipment with VR and CG creators, they are closer to end-users than professionals.
Unless affiliated with organizations providing abundant technical resources, VTubers primarily rely on consumer-level hardware and software rather than industrial-grade equipment \cite{Rohrbacher-2024-VTubing, ferreira-2022-concept}.
Moreover, during live streaming sessions, they are responsible for setting up equipment, performing as their avatars, and managing rendering processes independently.
To support their engagement performance, researchers have proposed systems to enhance the expressiveness, accessibility, and interactivity of VTubing \cite{Yeh-2022-CatEars, Shirai-2019-Reality, Tang-2024-Warudo, Li-2021-AliMe, Amato-2024-KawAIi, Li-2023-Blibug, Josyula-2024-Tactile, Zhu-2024-power, Chen-2024-Conan's}.
Tang et al. developed a system that triggers gestures and facial animations based on VTubers' expressions and voices, helping them maintain their virtual identity while reducing cognitive loads \cite{Tang-2021-AlterEcho}.
Additionally, solutions like REALITY \cite{Shirai-2019-Reality} and Warudo \cite{Tang-2024-Warudo} aim to ease motion capture and avatar control with more accessible VTubing technologies.

While novel system research for VTubing has been emerging, no empirical research provides a big picture of VTuber’s processes and experiences in practice to encourage follow-up research. 
While Jiang et al. and Ferreira et al. reported on representative motion capture tools for VTubing \cite{jiang-2023-better, ferreira-2022-concept}, their focus was on specific equipment rather than providing a comprehensive overview of available options. 
To fill this gap, we conducted desk research and interviews with professional VTubers to gain deeper insights into the equipment they use and the challenges they face at each step of the VTubing process.
Within the broader research context, our study contributes to the understanding of VTubing tools as well as consumer-grade tools for CG and VR content creation by examining how non-professional users endeavor to create and manage high-quality avatar content.

\begin{figure*}
    \centering
    \includegraphics[width=\linewidth]{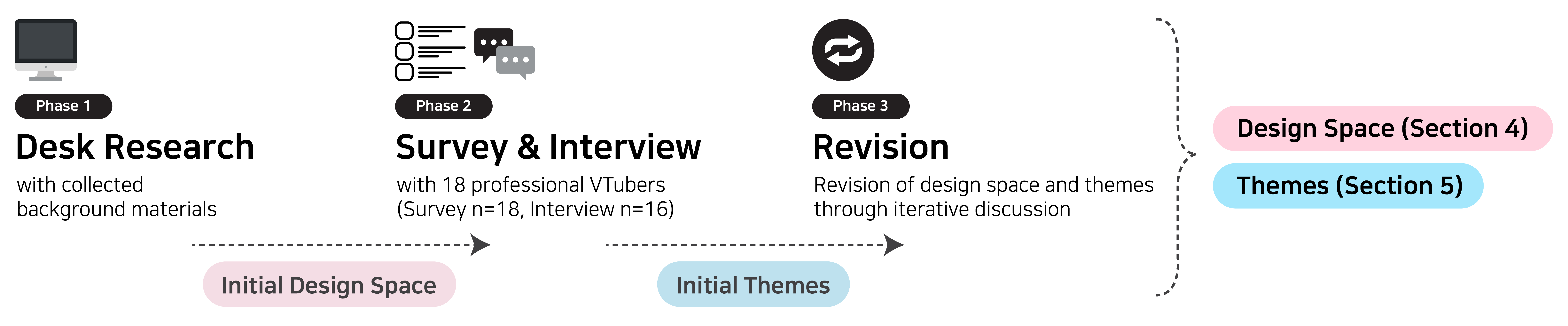}
    \caption{Overall research flow of this study included: 1) desk research involving material collection from online resources, and 2) preliminary surveys and interviews with 18 professional VTubers. The initial design space and themes identified through these processes were refined and finalized through iterative discussions and revisions. Findings are detailed in Sections \ref{sec:designspace} and \ref{sec:interview}.}
    \Description{The image illustrates the three phases of the study design, summarized as follows:
    Phase 1: Desk Research Represented by an icon of a desktop computer, this phase involves collecting background materials from online resources to construct the initial design space.
    Phase 2: Survey \& Interview Depicted with icons of checklists and speech bubbles, this phase involves conducting surveys with 18 professional VTubers (n=18) and interviews with 16 of them (n=16) to gather insights. Outcomes include the identification of the ``Initial Design Space'' and ``Initial Theme.''
    Phase 3: Revision Represented by a circular arrow icon, this phase involves refining the design space and themes through iterative discussions and revisions.
    At the bottom right, the image links the refined outcomes to specific sections of the study: ``Design Space'' (detailed in Section 4) and ``Theme'' (detailed in Section 5).}
    \label{fig:method}
\end{figure*}

\section{Study Design}
To address RQs, we collected background material from online resources and interviewed 16 among 18 professional VTubers who participated in the preliminary survey (Figure \ref{fig:method}).

\subsection{Background Material Collection}
Before designing the interview protocol, we first collected background material from various online sources to deepen our understanding of the equipment and processes commonly used by VTubers.
Between June and July 2023, we searched for and reviewed YouTube videos and forum posts using keywords like “How to start VTubing,” “VTuber model types,” and “VTuber co-streaming.”
These resources offered guidance for novice VTubers on preparation for debut, tutorials on avatar model types and creation methods, and instructions for setting up avatar control systems.
We compiled information on the hardware and software mentioned, including features, pros and cons, and pricing.
To verify specifications and capabilities, we also consulted the official websites of manufacturers and developers.
Additionally, we tested some available equipment to gain a firsthand understanding of its functionality.

Once we had compiled a list of equipment, we compared it to the live-streaming equipment design space proposed by Drosos and Guo \cite{Drosos-2022-DesignSpace} to identify overlaps between VTuber and real-person streamer equipment.
We categorize the VTuber-specific equipment under codes such as ``Model Type,'' ``Model Creation'', and ``Motion Tracking.''
The lead author then established the initial dimensions of the VTuber-specific equipment design space.

Throughout this process, we noted that most VTubers stream using desktop setups, integrating various devices and software to control their avatars.
To focus on general VTubing experiences, we excluded VTubing mobile apps, such as REALITY \cite{Shirai-2019-Reality}, from our research scope.

\subsection{Participant Recruitment}
\begin{table*}[]
    \centering
    \begin{tabular}{cccccccccc}
        \toprule
         ID & Main Platform & Region & \makecell{Gender\\(VTuber/Avatar)} & Model Type(s) & \makecell{Affiliated\\with an Agency} & \makecell{Number of \\Subscribers} & \makecell{VTuber EXP\\(Years)} \\
         \toprule
         P1 & Twitch & Republic of Korea & M/M & 3D & Yes & 165K & 5 \\
         \midrule
         P2 & Youtube & Republic of Korea & M/F & 2D & No & 58K & 1.5 \\
         \midrule
         P3 & Twitch & Republic of Korea & F/F & 2D$^*$, 3D & Yes & 8K & 1\\
         \midrule
        P4 & Youtube & Republic of Korea & F/F & 3D & No & 72K & 1\\
        \midrule
         P5 & Twitch & Republic of Korea & F/F & 2D, 3D$^*$ & No & 51K & 1\\
         \midrule
         P6 & Twitch & Republic of Korea & F/F & 2D & No & 83K & 5  \\
         \midrule
          P7 &  Twitch & Republic of Korea & F/F & 2D, 3D$^*$ & No & 6K & 3\\
          \midrule
         P8 & Twitch & Republic of Korea & F/F & 3D & No & 15K & 3  \\
         \midrule
         P9 & Twitch & Republic of Korea & F/F & 3D & No & 6K & 1 \\
         \midrule
         P10 & AfreecaTV & Republic of Korea & F/F & 3D & No & 12K & 1  \\
         \midrule
         P11 & Twitch & USA & M/M & 2D & No & 13K & 1.5  \\
         \midrule
         P12 & Youtube & USA & M/M & 2D & No & 56K & 2 \\
         \midrule
         P13 & Twitch & USA & M/M & 2D & No & 60K & 2.5  \\
         \midrule
         P14 & Twitch & Philippines & M/M & 2D & No & 20K & 1.5  \\
         \midrule
         P15 & Twitch & USA & F/F & 2D & No & 92K & 2  \\
         \midrule
         P16 & Youtube & Japan & F/F & 2D & No & 75K & 3  \\
         \midrule
         S1 & Twitch & Canada & M/M & 2D & No & 6K & 1.5  \\
         \midrule
         S2 & Twitch & Republic of Korea & F/F & 2D & No & 47K & 1  \\
         \bottomrule
    \end{tabular}
    \caption{
        Information (as of 1 September 2023) about VTubers who participated in the interview.
        The symbol $^*$ indicates the model type that is primarily in use.
    }
    \Description{Table 1 provides information about VTubers who participated in interviews as of 1 September 2023. The table contains the following columns: ID, Main Platform, Region, Gender (VTuber/Avatar), Model Type(s), Affiliated with an Agency, Number of Subscribers, and VTuber EXP (Years). Summary of key data includes the following: Most VTubers use either 2D or 3D avatars, with some using both; Subscribers range from as low as 6K to as high as 165K; VTuber experience spans from 1 to 5 years; A few VTubers are affiliated with agencies, while the majority are independent.}
    \label{tab:participant}
\end{table*}
Based on the insights from the background material analysis, we conducted a preliminary survey and semi-structured interviews with VTubers. 
To ensure diverse representation, we compiled a list of potential participants by searching VTuber wikis \cite{VTuberWiki} and YouTube, creating a pool of 239 VTubers with a variety of genders, subscriber counts, agency affiliations, avatar model types, cultural backgrounds, and streaming platforms.
We contacted 149 VTubers whose email addresses or Discord accounts were publicly available to invite them to participate in the survey and interview.
For VTubers affiliated with agencies that restrict direct contact, we approached relevant stakeholders within those agencies.
Ultimately, out of 149 VTubers, 25 responded: 16 agreed to interviews, 2 participated only in the survey, and 7 declined.

During the recruitment process, we observed that VTubers were generally cautious about sharing personal information.
Female VTubers and those using 2D models were more prevalent than male VTubers and 3D model users, consistent with market trends and prior research \cite{BRI-VTuberMarket,Wan-2024-Investigating,Lu-2021-MoreKawaii}.
To reduce potential bias, we focused on achieving a balanced sample rather than increasing the sample size arbitrarily.
Qualitative research literature suggests that meaningful insights can often be obtained from relatively small sample sizes, typically between 9 and 17 participants \cite{hennink-2022-sample}.
Additionally, we observed that no new themes or insights emerged after the 13th interview, indicating that data saturation had been reached.
Based on these observations and the need to balance participant diversity with depth of analysis, we determined that a sample size of 16 was appropriate for the scope of this study and concluded the recruitment process.
This decision ensured that our sample provided sufficient depth and diversity to address our research objectives without compromising the quality and manageability of the analysis.

\subsection{Interviewee Backgrounds}
Sixteen VTubers (P1-P16) participated in both the survey and interviews, while two others (S1-S2) participated only in the survey. 
Table \ref{tab:participant} provides a summary of information about our participants.
The participants hailed from diverse cultural backgrounds, including Republic of Korea, the USA, the Philippines, Japan, and Canada.
Their subscriber counts ranged from 6K to 165K, with experience levels varying from 1 to 5 years.
Most participants streamed primarily on Twitch, with a few active on YouTube and AfreecaTV.
Eight participants used 2D avatars, five used 3D avatars, and three used both.
The participants consisted of 10 females and eight males, with all of them except P2 using avatars that matched their biological gender.
Most participants were independent VTubers, while P1 and P3 were affiliated with VTuber agencies.

\subsection{Interview Protocol}
The survey gathered demographic data, as well as information on their streams, avatar models, methods and costs of avatar creation, VTubing equipment, and the extent of avatar control.
The survey took approximately 20 minutes to complete.
Following the survey, we conducted semi-structured interviews with 16 participants to delve deeper into their VTubing experiences.
The interviews, conducted via Discord between August and October 2023, explored four key areas of the VTubing process:
\begin{enumerate}
    \item Experience in designing and creating avatars, including key considerations and challenges
    \item Setup of VTubing equipment for avatar operation, including variations based on content
    \item Avatar control and interaction with viewers and other streamers during streams
    \item Efforts to maintain the VTuber identity and communicate with viewers beyond streams
\end{enumerate}
Additionally, participants were asked to compare the advantages and limitations of VTubing with real-person streaming.
To ensure the relevance of our questions, we tailored them to each participant’s content and avatar setup based on their survey responses and YouTube channels.
Although conducted remotely, participants provided detailed explanations of their avatar control processes by sharing videos, photos, or screen shares to demonstrate how they operated their software.

Interviews were conducted in Korean, English, and Japanese.
Native Korean and fluent English-speaking authors facilitated the session, while an external translator was present for Japanese interviews.
Researchers took notes and recorded all sessions using OBS \cite{OBS} with the participants' consent.
Each session lasted 1.5 to 2 hours, and participants were compensated with USD 40.
If further clarification was needed, we conducted follow-up interviews lasting 30 minutes to an hour, offering the same compensation.

\subsection{Data Analysis}
To analyze the interview, we transcribed all recorded sessions using Clovanote \cite{Clovanote} and translated non-Korean interviews into Korean.
We employed a two-phase process using an open coding method \cite{Strauss-1997-GroundedTheory}.
In the first phase, the lead author performed open coding on a subset of the transcripts, following a qualitative thematic analysis approach \cite{Virginia-2006-Qualitative}.
The initial codes focused on VTubing experiences, challenges, and strategies for designing avatars, setting up equipment, and controlling avatars.
Overarching themes were derived from these codes.
In the second phase, co-authors independently coded the transcribed interviews.
After completing the coding, the authors discussed any discrepancies and reached a consensus.
The themes were revised to align with the agreed-upon codes and organized into sub-themes.
Finally, we reviewed the consistency of the themes and sub-themes, which were mapped to key areas: avatar design, creation, setup, control, and interaction (presented in Section \ref{sec:interview}).

Additionally, while analyzing the interviews, we continuously refined the dimensions of the initial design space using an iterative process that combined inductive and deductive methods.
We mapped the participants' equipment setups onto the design space dimensions to ensure they adequately reflect their experiences and equipment.
To further validate the design space, we compared our dimensions with those proposed by Drosos and Guo \cite{Drosos-2022-DesignSpace}, ensuring consistency in categorization and fidelity format.
This process was repeated until all authors reached a consensus.

\begin{figure*}
    \centering
    \includegraphics[width=\linewidth]{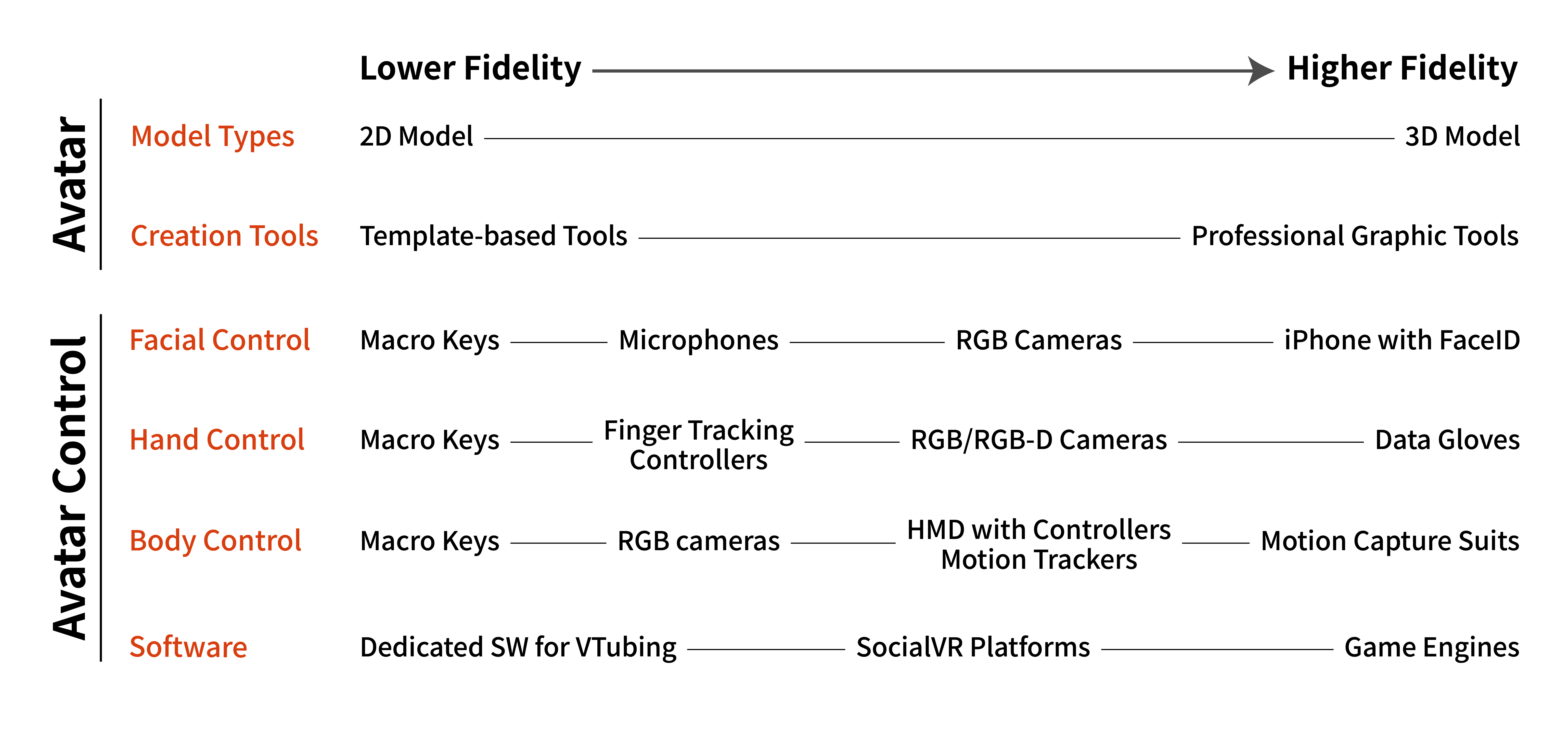}
    \caption{We formulated six dimensions, in addition to the existing live-streaming design space \cite{Drosos-2022-DesignSpace}, to include equipment choices related to virtual avatars.}
    \Description{The figure presents six dimensions relating to VTuber streaming equipment's fidelity and avatar control mechanisms, expanding upon existing live-streaming design spaces. The dimensions are divided into two main categories: “Avatar” and “Avatar Control.” A left-to-right arrow at the top indicates increasing fidelity, ranging from lower to higher. The following is the description of the Avatar group: Model Types: Ranges from 2D Models (lower fidelity) to 3D Models (higher fidelity); Creation Tools: From template-based tools (lower fidelity) to professional graphic tools (higher fidelity). The following is the description of the Avatar Control group: Facial Control: Progresses from macro keys (lower fidelity) to microphones, RGB cameras, and iPhones with FaceID (higher fidelity); Hand Control: Begins with macro keys, advancing to finger-tracking controllers, RGB/RGB-D cameras, and ultimately data gloves (higher fidelity); Body Control: Moves from macro keys to RGB cameras, head-mounted displays (HMD) with controllers, motion trackers, and finally motion capture suits (highest fidelity); Software: Starts with dedicated software for VTubing (lower fidelity), advancing to social VR platforms, and game engines (higher fidelity). Each dimension offers a continuum of fidelity for tools and technologies involved in VTuber live-streaming, reflecting the complexity and sophistication of the equipment in use.}
    \label{fig:design-space}
\end{figure*}

\section{The Design Space of VTubing Equipment (RQ1)}
\label{sec:designspace}
This section outlines the findings related to the key dimensions of the design space for VTubing equipment, as identified through the analysis.
As VTubing is founded upon the principle of live streaming, the VTubing equipment essentially includes the hardware and software equipment that real-person live streamers use. 
Accordingly, we define the design space for VTubing by extending that of conventional live streaming, as established by Drosos and Guo \cite{Drosos-2022-DesignSpace}, to include additional dimensions. 
What differentiates VTuber setups is the use of avatars for self-representation and the specialized equipment required to control these avatars, which introduces two key groups of dimensions: Avatar and Avatar Control. 
Figure \ref{fig:design-space} provides an overview of the extended design space for VTubing equipment. 
For further information on live-streaming-related dimensions, please refer to the paper by Drosos and Guo \cite{Drosos-2022-DesignSpace}.

\subsection{Avatar}
\label{subsec:designspace-avatar}
\begin{figure*}
    \centering
    \subfigure[Example of 2D Model]{
        \includegraphics[width=.45\linewidth]{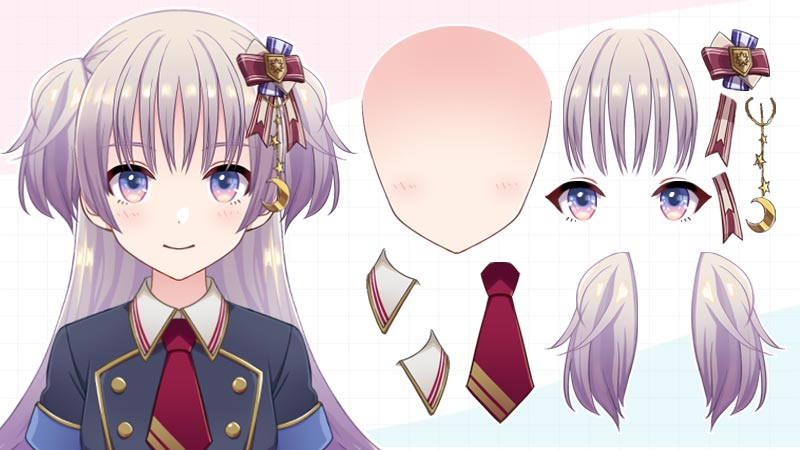}
        \label{fig:2d-model}
    }
    \subfigure[Example of 3D Model]{
        \includegraphics[width=.45\linewidth]{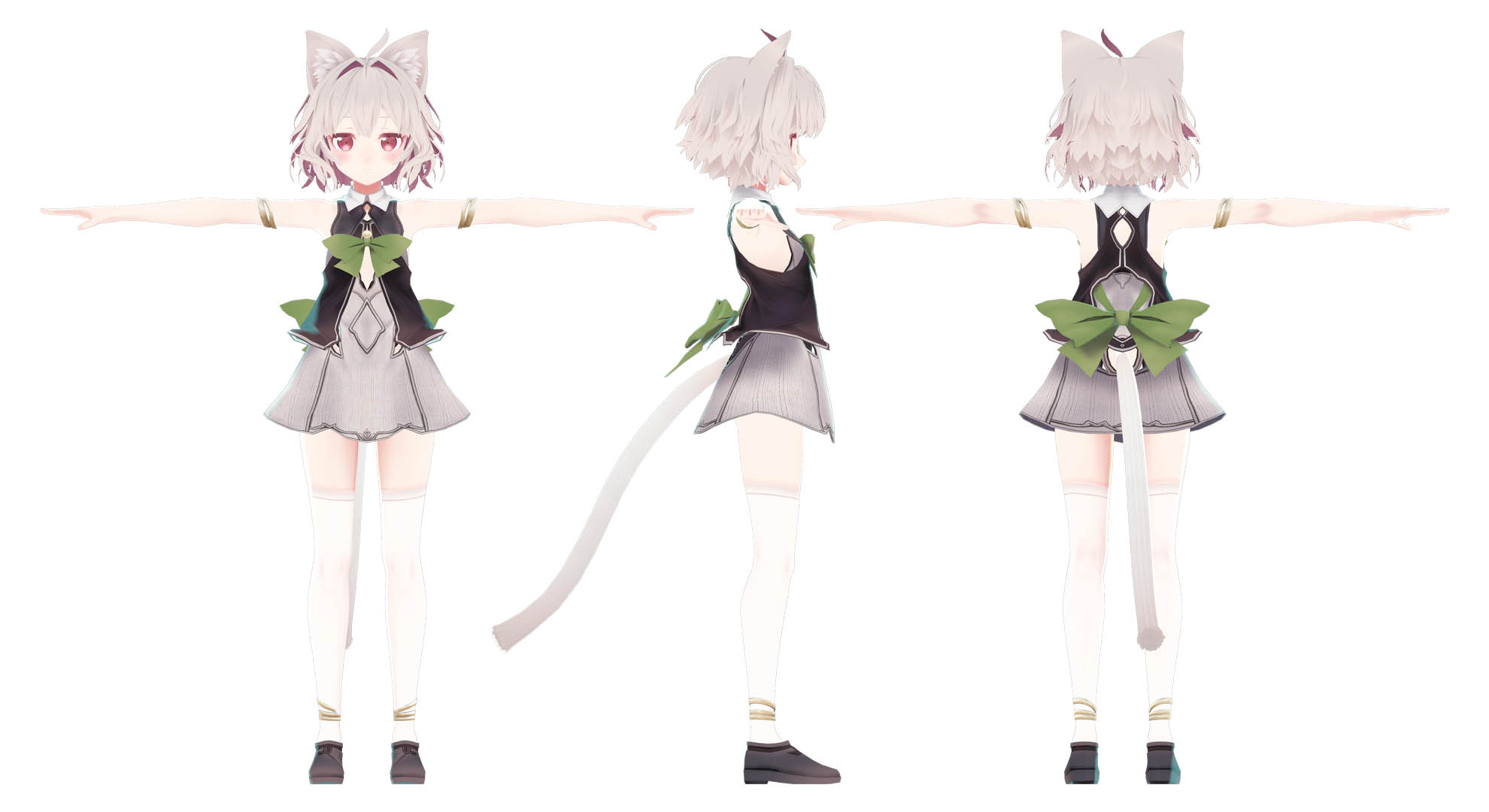}
        \label{fig:3d-model}
    }
    \caption{Examples of VTuber avatar model types (Image retrieved from \cite{Model_2D, Model_3D})}
    \Description{This figure illustrates examples of VTuber avatar model types, showing both 2D and 3D avatars. The figure is divided into two subfigures. Subfigure (a): Example of 2D Model: On the left side of the figure is an example of a complete 2D model’s appearance. The right side of the figure shows a flat, illustrated avatar model divided into individual parts. The avatar has a detailed face with large expressive eyes, short silver hair with purple hues, and accessories such as ribbons and a school uniform tie. The separate pieces of the model (hair, face, ribbons, and other accessories) are displayed around the face, indicating how the 2D model is assembled for animation. Subfigure (b): Example of 3D Model: This depicts a 3D avatar, shown from three different angles: front, side, and back. The model is a humanoid character with cat-like features, including cat ears and a tail. The character is wearing a school-style outfit, and the 3D form shows more depth, shading, and structure, providing a sense of volume and physicality.}
    \label{fig:model-types}
\end{figure*}

\subsubsection{Model Types}
\label{subsubsec:designspace-modeltype}
Avatars are generally categorized into 2D and 3D models (Figure \ref{fig:model-types}), with 2D models being more widely used in the current VTubing industry \cite{BRI-VTuberMarket}.
Typically, VTubers use one type of avatar, although some may use both, choosing between them based on the stream content.

\textbf{2D Model (Figure \ref{fig:2d-model}):}
2D models consist of flat, two-dimensional illustrations.
To animate these illustrations in VTubing, two approaches are employed: playing sequential static images or animating using Live2D \cite{Live2D}.
The sequential image approach resembles cel animation, where a series of images are played in sequence to create the illusion of movement.
For example, VTubers might use two images of an avatar with open and closed lips and loop them based on whether they are speaking. 
While this approach is cost-effective, it does not appear realistic and limits the diversity of animations.

To achieve more realistic and vivid expressions, most 2D models use the Live2D format, with layers representing body parts, such as the eyes and mouth (Figure \ref{fig:2d-model}).
Each layer has an animation trajectory and deformation properties that enable appropriate animations based on user input.
VTubers can animate these layers through various input methods, such as button presses or facial tracking, allowing the avatar to express a wide range of movements.

However, the layer-based animation system of Live2D poses challenges in achieving smooth and fluid motion.
High-quality avatar animations require numerous layers, sophisticated animation trajectories, and deformation settings.
This labor-intensive process often restricts Live2D avatars to focusing primarily on facial expressions and upper body movements rather than full-body animation.
The complexity of this animation process also increases the cost of avatar creation, with prices ranging from \$2,000 to \$5,000, and higher for more detailed and complex animations.

\textbf{3D Model (Figure \ref{fig:3d-model}):}
3D models are commonly used in game engines, social VR platforms, and other virtual environments.
3D models incorporate a skeletal structure that defines body joints and blendshapes to control facial expressions, similar to the conventions used for humanoid avatars in computer graphics \cite{Baran-2007-Automatic, Lewis-2014-Blendshape}.
VTuber 3D models often include additional blendshapes for more creative, anime-style expressions, such as modifying the iris or adding blush effects.

By manipulating the skeletal structure and blendshape coefficients of their 3D models through various input devices, VTubers can achieve full-body and facial animations.
Thanks to their compatibility with various virtual environments, 3D models allow VTubers to explore creative possibilities across diverse virtual spaces.
However, these models require high-performance hardware for optimal operation and are costly to commission. 
The cost of creating a custom 3D avatar from a professional artist ranges from \$500 to \$15,000, depending on the desired level of detail.

\subsubsection{Creation Tools}
\label{subsubsec:designspace-creationtool}
VTubers can create avatars using software with predefined templates or professional graphic tools for custom designs from scratch.

\textbf{Template-based Tools:}
Template-based tools let users create 3D models by assembling avatars from predefined body templates and parts.
VRoid Studio \cite{VRoidstudio,Isozaki-2021-VRoidStudio} is a popular choice for creating anime-style avatars, while MetaHuman \cite{MetaHuman} is often used for generating more realistic avatars.
These tools provide intuitive interfaces, enabling non-experts to select and customize different body parts and styles conveniently.
However, because these tools rely on predefined templates with limited customization options, avatars often have similar appearances, making them less distinct and unique.

\textbf{Professional Graphics Tools:}
These tools allow users to create avatars from scratch, providing complete creative control.
The software choice depends on the avatar type.
For 2D models, programs like Adobe Photoshop \cite{Photoshop} and Clip Studio \cite{Clipstudio} are commonly used for illustration, while Live2D Cubism \cite{Live2D} is employed to animate the illustrations.
For 3D models, industry-standard tools such as Blender \cite{Blender} are used for 3D modeling, with Adobe Photoshop \cite{Photoshop} often used for authoring textures.
While these tools enable the creation of highly customized and unique avatars, they require specialized skills, making it challenging for non-professionals to achieve their desired outcomes without substantial time and effort.

\subsection{Avatar Control}
\label{subsec:designspace-avtarcontrol}
Controlling avatars in VTubing requires devices to capture facial, body, and hand movements, with software mapping them to those of the avatar.
VTuber can use various hardware, ranging from macro keys that trigger predefined animations to high-fidelity tracking devices for accurate motion capture.
The extent of controllable body parts and hardware fidelity depends on the capabilities of the operation software.

\subsubsection{Common Control Devices for All Body Parts}
\label{subsubsec:designspace-common}
Different devices are needed to control individual body parts, but macro keys and RGB cameras can be used for overall body control.

\textbf{Macro Keys:}
Macro keys are the lowest fidelity control device, triggering predefined animations for all body parts.
They are useful when pose-tracking hardware is unavailable or for triggering expressive anime-style expressions such as blushing cheeks that pose-tracking hardware cannot express.
The most basic hardware is a keyboard, where unused keys can be assigned to trigger animations.
However, using a keyboard can lead to accidental triggers when mapped keys conflict with other software.
To prevent key conflicts and assign more animations, VTubers often use dedicated macro devices such as the Elagto stream deck \cite{ElgatoStreamDeck} and Elgato stream deck pedal \cite{ElgatoStreamDeckPedal}.

\textbf{RGB Cameras:}
RGB cameras track the user's facial expressions and pose to control the avatar.
Various devices, from webcams to DSLRs, can be used for this purpose, with webcams being the most popular due to their affordability and ease of setup.
Positioned in front of the VTuber, the camera captures RGB images.
VTubing software then analyzes the facial expressions and poses using computer vision technology and applies the tracked poses to the avatar in real time.

\subsubsection{Facial Control}
\label{subsubsec:designspace-facialcontrol}
VTubers uses microphones and FaceID-ready iPhones in addition to the aforementioned common equipment for facial control.

\textbf{Microphones:}
Microphones capture the user's voice and translate it into corresponding lip movements on the avatar.
This method is beneficial when it is impossible to track facial expressions, such as when the user is wearing an HMD.

\textbf{iPhones with FaceID:}
Similar to RGB cameras, iPhones with FaceID are positioned in front of the VTuber to track facial expressions.
iPhones use a front-facing TrueDepth camera to capture a detailed depth map of the face \cite{FaceID}, enabling more precise tracking of facial expressions than RGB cameras.
This advanced tracking capability makes the iPhone a preferred choice among many VTubers for achieving sophisticated facial control.

\subsubsection{Hand Control}
\label{subsubsec:designspace-handcontrol}
VTubers use finger-tracking controllers, RGB-D cameras, and data gloves to control their avatar's hand movements.

\textbf{Finger-tracking Controllers:}
These are specialized VR controllers that detect each finger's movement based on grip sensors built into the controller but are not capable of capturing subtle motions.
They are commonly used in immersive virtual environments with HMDs to animate the avatar's fingers.
A well-known example is the Valve Index Controller \cite{ValveIndex}.

\textbf{RGB-D Cameras:}
RGB-D cameras function similarly to iPhones with facial tracking capabilities, utilizing depth maps to track hand poses.
These cameras are typically placed in front of an HMD or on a desk to monitor hand movements within their trackable range.
The Leap Motion \cite{LeapMotion} is widely used for this purpose because of its convenience, as allowing hand control without requiring additional wearable devices.
However, it has limitations, such as difficulty tracking fast hand movements or hands moving outside the camera's field of view, and it has a short detection range, restricting hand movements to areas near the camera.

\textbf{Data Gloves:}
Data gloves offer the highest fidelity in hand tracking by directly capturing detailed hand movements through sensors embedded in the gloves.
Devices like Xsens gloves by Manus \cite{XSensGlove} provide the most accurate hand control available.
However, due to their high cost and discomfort associated with wearing physical devices, they are only used by a few VTubers.

\subsubsection{Body Control}
\label{subsubsec:designspace-bodycontrol}
VTubers primarily utilize equipment specifically for body control to represent their full-body movements in 3D environments, such as social VR platforms or game engines.

\textbf{HMD with Controllers and Motion Trackers:}
HMDs, hand-held controllers, and motion trackers are widely used to generate articulated animations in VR environments.
When using only HMDs and their controllers, VTubers can control their avatar's upper body and head orientation.
To achieve full-body tracking, VTubers use additional motion trackers to track lower-body movements.
Famous products in this category include the Meta Quest \cite{MetaQuest} and the HTC Vive \cite{Vive}.

\textbf{Motion Capture Suits:}
Motion capture suits offer the highest fidelity in body control, with leading products from companies like XSens \cite{Xsens} and Perception Neuron \cite{PerceptionNeuron}.
These suits provide the most accurate full-body tracking but are expensive, limiting their use to a smaller subset of VTubers.
Additionally, some suits are susceptible to interference from magnetic fields, requiring adjustments to the VTubing space, such as modifying or relocating furniture and objects, to prevent disruptions \cite{Lee-2024-Wishbowl}.

\subsubsection{Software}
\label{subsubsec:designspace-software}
VTubers utilize specialized software with the hardware introduced earlier to operate their avatars effectively.
The final rendered images are captured and transferred to broadcasting software, such as OBS \cite{OBS}, for live streaming.

\textbf{Dedicated Software for VTubing:}
VTubers who stream in non-VR environments control their avatars using dedicated VTubing software.
The choice of software depends on the type of avatar model used.
VTubers using Live2D models typically use VTube Studio \cite{VTubestudio}, while those with 3D models often rely on VSeeFace \cite{VSeeFace}.
These solutions support integration with tracking devices such as RGB/RGB-D cameras and iPhones for facial and hand tracking.
Furthermore, these solutions also support macro keys to trigger animations, offering more diverse expression options.

\textbf{Social VR Platforms:}
Social VR platforms like VRChat \cite{VRChat} provide a virtual venue for VTubers to stream using 3D models.
In these environments, VTubers use equipment like HMDs, controllers, and motion trackers to move their avatars freely within a 3D space and interact with virtual objects.
These platforms also allow VTubers to stream from diverse perspectives using virtual cameras that can create dynamic visuals by presenting their avatars from multiple angles during live streams \cite{Freeman-2024-MyAudience}.

\textbf{Game Engines:}
VTubers or teams with expertise in 3D content creation often use game engines like Unity \cite{Unity} or Unreal Engine \cite{UnrealEngine} to develop custom content.
Game engines offer the flexibility to design complex virtual environments and support a wide range of high-fidelity devices, including HMDs, data gloves, and motion capture suits.
These tools allow for more immersive avatar movement, enhanced object interactions, and higher-quality graphics, captivating audiences with superior visuals.
By leveraging these capabilities, VTubers can create unique content, such as virtual music concerts, that deliver dynamic and high-quality presentations beyond what is typically achievable in social VR environments \cite{Lee-2023-Ju.Taime}.
\section{Current Practices and Experience of VTubers (RQ2)}
\label{sec:interview}
Based on the identified design space and interviews with professional VTubers, we comprehensively explored the current practices, strategies, experiences, and challenges of VTubers throughout the VTubing process. 
We organized the findings into the following stages: ideating and creating avatars, setting up equipment for VTubing, controlling avatars during VTubing, and interacting with others through their avatars.

\subsection{Avatar Ideation}
Avatar ideation involves designing the avatar's appearance and virtual identity before starting VTubing.
During this creative process, VTubers develop the look and overall concept of their avatars.
Conceptualizing an avatar from scratch often leads to creative struggles.
P1 described the process as follows:
\textit{``Ideating avatar is as challenging as trying to have a universally appealing appearance from birth,''}
highlighting the complexity and pressure during the ideation process.
Below, we elaborate on strategies used by participants to overcome challenges and design practices to attract viewers and sustain VTubing.

\subsubsection{Primary Strategy: Drawing from Themselves}
Many participants designed their avatars by drawing inspiration from their appearance, personality, and preferences.
P15 explained, \textit{``I think most avatars are based on something the creator can relate to. So, I always take a little from myself or my experiences and put them into my character.''}
P9 designed a fox avatar inspired by her appearance and personality, and P11 combined his experience as a bartender with his fondness for cats to create a cat bartender avatar.

This strategy was also influential in designing virtual identities.
Participants intentionally ensured consistency between their real and virtual identities, viewing this alignment as essential for sustainable VTubing.
While VTubers can adopt entirely new virtual identities, many found acting out a new persona challenging, especially without professional voice-acting experience.
P2 noted, \textit{``Unless you can completely separate yourself from your identity, acting as an entirely new persona requires significant effort. You must be constantly aware to ensure that your casual speech tones do not emerge so that you can maintain the new persona throughout the stream.''}
P11 also echoed, \textit{``I tried role-playing a different identity, but over long streams, my real personality would come through, making it hard to sustain the role.''}
Furthermore, VTubers emphasized the importance of entertaining viewers as entertainers, noting that portraying an entirely new identity could impede their ability to engage effectively.
P11 observed that maintaining a persona while providing entertainment often requires sacrificing one aspect if the two cannot be balanced.
Similarly, P2 explained that streaming as their authentic self enables them to deliver the highest level of energy during the stream.

Furthermore, VTubers emphasized that maintaining a different identity could negatively impact their well-being.
P3 shared a story about a friend who quit VTubing after experiencing significant stress from portraying a sexually emphasized identity that did not suit her, explaining that playing an identity misaligned with one's true self can be mentally burdensome.
P7 echoed, \textit{``Portraying a new identity is akin to wearing a mask. Pretending to be kind, tolerating teasing from viewers, and maintaining this facade was challenging for me. I wanted to be happy and enjoy streaming, but forcing myself to act in a way I disliked made me consider quitting, even if I succeeded.''}
P13 similarly emphasized that adopting an identity closely aligned with their own was essential for sustaining their VTubing career, stating, \textit{``I would get personally exhausted if I could not align my avatar with my true self. If I cannot overcome the burden of acting as a different identity, I don't think I will be able to enjoy the work any longer.''}

\subsubsection{Considerations: Harmonizing Voice and Appearance}
While participants designed their avatars based on their identities, they also emphasized the importance of harmonizing the avatar's appearance with their voice.
Participants noted that mismatches between an avatar's appearance and voice---such as a mature, sexy avatar paired with a cute, childlike voice, or vice versa---could cause discomfort for viewers.
To prevent such dissonance, VTubers often designed their avatars to align with their natural voices.
For example, P5 explained her decision to design a schoolgirl avatar: \textit{``When we see an anime character, we instinctively associate a particular voice with its image and feel discomfort when they don't match. I did not want to evoke that sense of discomfort. So, when I designed my avatar, I thought my voice fit best with a teenage girl rather than an adult or sultry woman. As a result, I designed a schoolgirl avatar.''}

P3, leveraging professional voice-acting experience, adopted a different approach first by designing her ideal avatar and then performing voice-acting to align with the avatar's appearance.
However, for most participants, this strategy proved challenging to sustain.
P6 pointed out the difficulty of voice-acting during extended streams, noting, \textit{``I tried raising my voice pitch to match my avatar's appearance, but it is exhausting when the stream goes longer. I typically stream for 8 hours, sometimes up to 15 hours, but If I keep my voice high, I can only manage about four hours. So now, I just speak comfortably.''}

\subsubsection{Developing over Time}
While many participants finalized their avatar's appearance and identity during the ideation phase, others refined them over time as they continued VTubing.
P10, who changed avatars five times, shared, \textit{``I did not solidify my identity until I met different people through streaming. At first, I used a generic character because I was unsure of what I wanted, but eventually, I developed current goofy identity.''}
However, some VTubers cautioned that this iterative approach can be risky for those planning to focus on role-playing or maintaining a consistent concept.
P1 noted, \textit{``If you plan to role-play, you need to establish your identity from the start by the time you debut.
With the established identity, you should cultivate a fan base of those who resonate with and support your identity.''}
Similarly, P3 emphasized the importance of careful initial planning, noting that changing the avatar later on can be challenging.
\subsection{Avatar Creation}
Creating an avatar is essential in VTubing, embodying the VTuber’s imagination.
However, most VTubers lack the professional design skills to achieve their desired look.
While some attempt to use template-based tools to create avatars, these tools provide limited customization, resulting in models that do not meet their desired quality.
As a result, most VTubers turn to professional artists for higher-quality avatars.
Within the budget constraints, VTubers adopt various strategies to secure the best possible avatar.

\subsubsection{Creation with Template-based Tools}
Some VTubers attempted to create their avatars using template-based tools like VRoid Studio \cite{VRoidstudio, Isozaki-2021-VRoidStudio}.
However, only two participants used VRoid avatars, and both eventually switched to higher-quality models after temporary use.
Participants commonly noted that template-based avatars looked generic and lacked detailed body features.
P1 remarked, \textit{``All VRoid avatars look the same since the customization options for the body, face, eyes, nose, and mouth are limited.
For example, creating hair using VRoid is like putting a few large strands of seaweed on the head, which made it impossible to create the detailed hairstyle I desired.''}
These limitations made it difficult for participants to achieve their desired aesthetic or convert 2D models to 3D while maintaining their unique style.
Consequently, most preferred to hire professional artists to create original avatars.

\subsubsection{Outsourcing to Experts}
For outsourcing, VTubers searched for artists through social media and freelancer platforms such as X and Fiverr \cite{Fiverr}.
After reviewing the portfolios of artists, VTubers chose artists whose work was similar to the look they envisioned for their avatars.
After contacting an artist, VTubers provided visual references from games, anime characters, or online platforms like Pinterest \cite{Pinterest} in an effort to better describe imagined avatars in mind.
For instance, P5 shared visual references with detailed explanations, including the referenced avatar's personality, traits, and atmosphere, to improve the artist's understanding.
On the other hand, some VTubers provided only a simple description of their desired visuals and worked closely with the artist throughout the process.
P1 explained, \textit{``Since most of the visual work is done by the artist, I just requested only simple key points, like adding shark fins on the head, making the eyes blue, giving the teeth a shark-like appearance, and designing the outfit as a hoodie.''}

\subsubsection{Trade-off between Avatar Quality and Budget}
The cost of commissioning an avatar from a professional artist depends on several factors, including the scope and detail of the rigging and additional features like hair, outfits, and facial expressions.
P3 remarked, \textit{``If your character looks off, it might be due to a lack of budget. The more you invest, the more natural and human-like the movements will be.''}
Therefore, participants had to balance their budget with the desired quality of the avatar.

With budget constraints, most participants prioritized visual quality over rigging.
P16 noted, \textit{``VTubers use their avatar illustrations in many places to promote themselves. In many cases, these illustrations are displayed alongside those of other VTubers, so if the illustration is not appealing, it will not attract much attention.''}
However, some participants warned that compromising rigging quality too much could negatively affect the viewing experience.
P8 stated, \textit{``If the jaw moves awkwardly during a stream, it could result in a poor viewer experience and the viewers leaving the stream.''}
We found an interesting common strategy: many participants saved money by focusing their spending on key parts of their avatar, such as the face or upper body, which are often visible in streams, and skipping or lowering the quality of accessories, the lower body, or other less visible elements. 
Moreover, since creating an avatar from scratch can be costly, all participants who use original 3D avatars, except for P1 and P5, opted for kitbashing, a strategy discussed in the next section.

\subsubsection{Kitbashing: Mixing Multiple Ready-made 3D Avatars}
\label{subsubsec:avatar-creation-kitbashing}
After purchasing high-quality 3D models from asset marketplaces such as Booth \cite{Booth}, kitbashing is the process of mixing them up by combining and modifying different body parts, faces, and outfits to create a unique avatar. 
Participants either performed kitbashing using software like Blender \cite{Blender} or Unity \cite{Unity} or outsourced the task to professional artists.
Kitbashing is popular among VTubers as a cost-effective alternative to custom avatars, providing higher-quality results than template-based tools.
Most of the VTubers with 3D models in this study used kitbashed avatars.
P8 likened the process to creating a chimera, saying, \textit{``Much like creating a chimera, choosing the best parts from different models: taking the eyebrows from one model and the legs from another.''}

However, since kitbashing leverages existing assets, it shares the potential limitation of template-based tools; avatars may resemble those of other VTubers built on the same models.
To create more unique visuals, participants often performed ``facial surgery,'' which involves adjusting facial rigging to alter expressions, such as changing the shape of the eyes or mouth.
P7 noted, \textit{``If you do not perform facial surgery, people will say you look exactly like another VTuber since other VTubers use the same base models for kitbashing.''}
P10 added, \textit{``Since so many people use the same models for kitbashing, I felt the need to customize mine through surgery to make it unique.''}

While kitbashing offers greater customization at an affordable cost, it comes with several challenges.
Participants had to search through numerous marketplace assets to find compatible items.
P7 explained, \textit{``Each avatar has a different body shape, so you must buy clothing that fits your avatar. If the outfit does not match the avatar, it is difficult to make it work.''}
Therefore, many participants used popular models with broader compatibility, and sellers on the marketplace often provided compatibility lists for their assets.
Participants also tried to modify incompatible assets to fit their avatars if they aligned with the envisioned design.
However, assembling and modifying models could damage the rigging, leading to issues with avatar movement, and participants sometimes could not achieve their ideal look entirely through kitbashing.
P3 shared, \textit{``I initially wanted to create an original model but struggled to find a suitable artist, so I attempted kitbashing. However, the results fell short of my expectations. Since I had a clear vision of my character, I was unwilling to compromise on any aspect. For individuals like me with a strong, uncompromising vision for their avatars, results produced in low-flexible environments can be unsatisfactory.''}

\subsection{Avatar Setup}
\label{subsec:avatar-setup}
To operate their avatars in live streams, VTubers configured various software and hardware.
The setup varied depending on the avatar model type and display layout.

\subsubsection{Upper-body Setup}
\label{subsubsec:avatar-setup-upperbody}

\begin{figure}
    \centering
    \includegraphics[width=.29\linewidth]{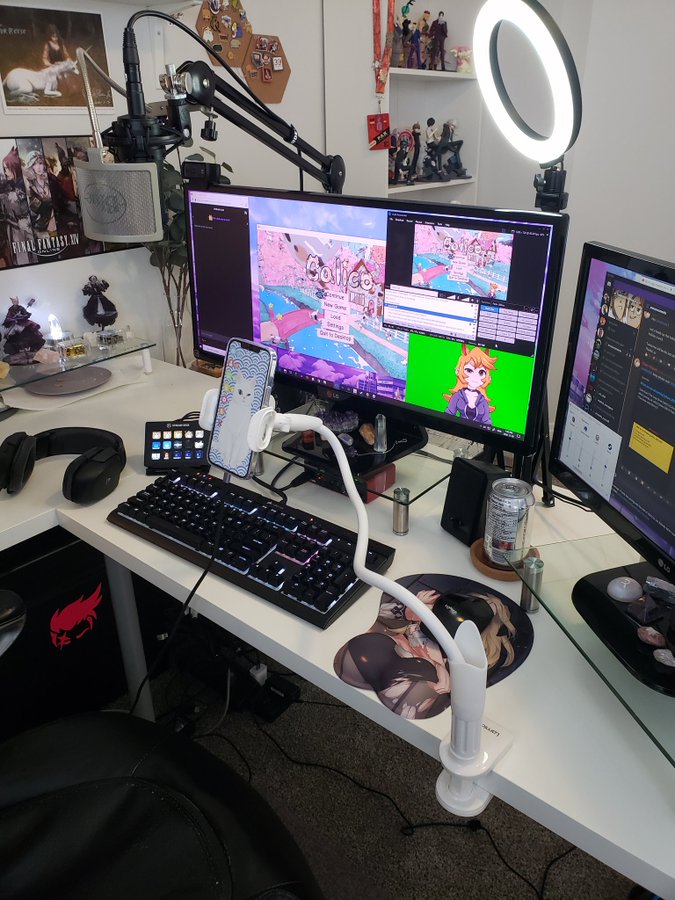}
    \includegraphics[width=.685\linewidth]{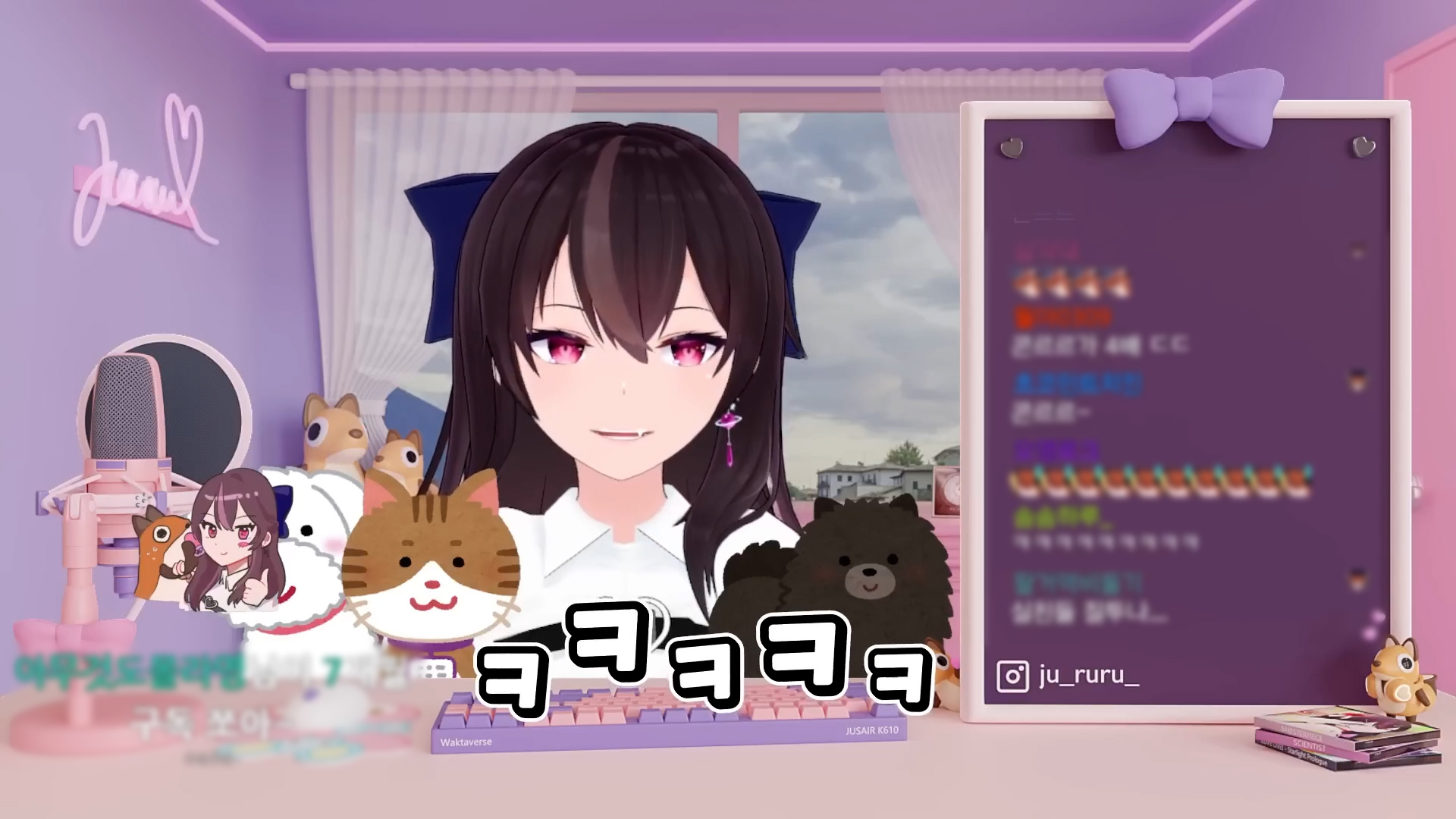}
    \caption{Examples of equipment setup and stream layout in upper-body setup (Image retrieved from \cite{Upperbody-setup, Upperbody-layout})}
    \Description{This figure shows examples of VTuber equipment setup and the stream layout used in an upper-body setup. The figure is divided into two parts: Left Side (Equipment Setup): This image showcases a VTuber's streaming desk setup. It features a large desktop monitor displaying a streaming software interface and additional overlays. The desk has several key components: A microphone mounted on a boom arm for capturing high-quality audio. A ring light is positioned above the monitor, which enhances the lighting for facial tracking. A gaming keyboard, mouse, and StreamDeck are placed to control their stream. The overall desk is organized for multi-tasking, with streaming and chat management tools visible on the screen. Right Side (Stream Layout): This image shows a sample VTuber stream layout from the viewer's perspective. The virtual avatar, a black-haired character, is centrally displayed and interacts with the audience. The background features a cozy room with a window view. On the right side of the screen is a live chat window, where viewer messages, represented by colorful emoticons, are displayed.}
    \label{fig:upper-body-setup}
\end{figure}

VTubers widely use the upper-body setup for both 2D and 3D models.
In this setup, as illustrated in Figure \ref{fig:upper-body-setup}, the upper half of an avatar is displayed in a corner or the center of the streaming screen, typically during chat or game streams.
All participants employed the upper-body setup in their streams, utilizing dedicated VTubing software and facial and hand control devices to animate their avatars.

Most VTubers in our study preferred using iPhones with FaceID for facial expression control, which provides high-fidelity tracking.
P6 noted, \textit{``I chose to use an iPhone with FaceID for facial tracking because it captures more detailed and varied expressions than a webcam.''}
Some VTubers further enhance facial animations by using additional software, such as VBridger \cite{VBridger}, which specializes in fine-tuning facial parameters. 
In addition, VTubers simultaneously use macro keys to trigger a range of expressions that facial tracking alone could not capture to increase expressiveness.

In contrast to facial control, only a few VTubers controlled their avatars' hands during the stream.
They used RGB or RGB-D cameras, such as webcams and Leap Motion, to capture their hands movements.
P1 was the only participant to employ a data glove for high-fidelity hand tracking.
Most VTubing software supports facial tracking, but hand tracking support varies.
If the software did not support their hand-tracking devices, VTubers had to integrate external motion-tracking software, like Webcam Motion Capture \cite{WebcamMocap}, to track hand movements and relay the data to the VTubing software.
Although this integration enabled VTubers to use various devices, it also required managing multiple software applications and handling a complex setup to ensure smooth avatar control.

\subsubsection{Full-body Setup}
\label{subsubsec:avatar-setup-fullbody}
\begin{figure}
    \centering
    \includegraphics[width=\linewidth]{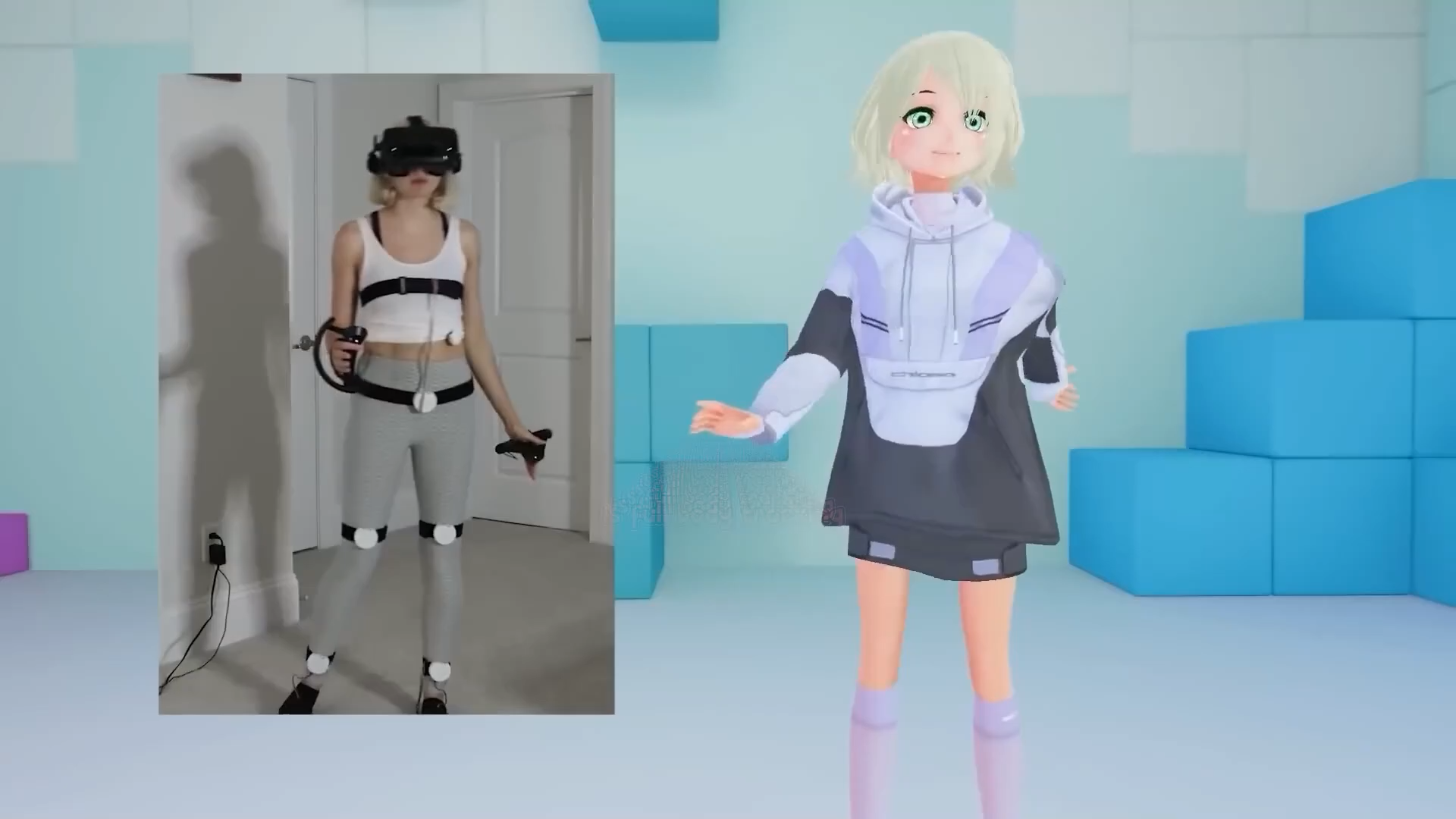}
    \caption{Examples of equipment setup and stream layout in full-body setup (Image retrieved from \cite{Fullbody-setup})}
    \Description{This figure shows examples of equipment setup and stream layout used for a full-body VTuber setup. The figure is divided into two parts: Left Side (Equipment Setup): This image features a person wearing equipment to control the avatar. The individual is equipped with a head-mounted VR display (HMD), motion trackers attached to their wrists, waist, knees, and ankles, and hand-held controllers in each hand. Right Side (Stream Layout): This shows the corresponding virtual avatar displayed on the stream. The avatar is a humanoid character with a youthful appearance, wearing a casual outfit with a hoodie and shorts. The avatar mimics the user's full-body movements in real-time, reflecting the position of arms, legs, and body posture. The background is a virtual 3D space with blue block-like structures.
    }
    \label{fig:full-body-setup}
\end{figure}

The full-body setup is used by VTubers with 3D models, allowing them to showcase their entire avatar in a 3D space during streaming (Figure \ref{fig:full-body-setup}).
This setup requires 3D-supported VTubing software, social VR platforms, or game engines.
All participants with 3D models except P10 have experience with the full-body setup.
P1 and P8 primarily used this setup for their streams, while others occasionally switched to it depending on the content.

Motion capture suits facilitate the highest fidelity tracking, but not all participants purchased them due to their high cost. 
VTubers typically used HMDs for head tracking, controllers for hand tracking, and motion trackers to track additional points such as waist and feet (Figure \ref{fig:full-body-setup}). 
P1 and P8 used finger-tracking controllers to generate sophisticated finger animations. 
In addition, similar to macro keys, VTubers often map expressive animations to the buttons on VR controllers, allowing them to overcome physical limitations and express distinctive reactions at the right moment during the stream. 
Wearing an HMD prevents the iPhone's face-tracking feature from working properly. 
As a remedy, VTubers rely on microphones to capture their speech and synchronize the avatar's lip movements with their voice. 
Participants often preferred devices from the same manufacturer for easier setup. 
However, P4 and P9 opted for a MixedVR setup, combining HMDs and motion trackers from different manufacturers to make use of their existing equipment. 
A typical example is the combination of a Meta Quest HMD with HTC Vive motion trackers, as Meta does not manufacture its own motion trackers. 
P9 noted the drawbacks of MixedVR setups, such as potential connection issues between devices leading to tracking errors and a more complex setup process.
\subsection{Avatar Control}
VTubers aim to make their avatars appear as lively as possible.
To achieve this, they try to exaggerate facial expressions and movements and trigger dynamic expressions in a timely manner.
This section presents the strategies VTubers employ to control various parts of their avatars and the challenges they encounter.

\subsubsection{Facial Control}
\label{subsec:facial-control}
P9 highlighted that \textit{``Even with the same avatar, the expressiveness can vary significantly depending on the person controlling it,''} emphasizing VTubers' efforts to enhance their avatars’ expressiveness.
VTubers primarily exaggerated their facial expressions beyond natural levels, ensuring that the facial tracking devices accurately captured their intended emotions.
They also typically used macro keys to trigger dramatic or anime-like expressions, such as turning the avatar's eyes into hearts, to entertain audiences and increase immersion.
P1 noted that such unrealistic expressions are a unique advantage of VTubers: \textit{``This is one of the distinctions. While it is difficult for a real person to exaggerate expressions like an anime character, a virtual avatar can do it easily. This characteristic allows VTubers to deliver more dramatic moods and expressions, thereby entertaining viewers.''}

However, managing multiple tasks during VTubing---running the content, interacting with viewers, and controlling the avatar---made it challenging to trigger expressions using macro keys in a timely manner.
VTubers should practice repeatedly to develop their reflexes to deal with this.
P4 likened their practice to that of an idol, saying, \textit{``A delay in the reaction can lead viewers to perceive them as unnatural, potentially breaking their immersion. To address this, I repeatedly practiced which button to press for which expression in various situations, like idols on the stage.''}
Despite the preparation, participants found it difficult to control their avatar's expressions during activities that required intense focus, such as gaming.
P1 explained, \textit{``Both hands are busy with the game, so it is hard to press a button to change expressions,''} highlighting the multitasking challenges in such scenarios.
Due to these difficulties, some VTubers chose to forgo using macro keys entirely.
P6 noted, \textit{``Since I usually stream games, I rarely trigger expressions with buttons, except when reacting to donations or chatting casually with viewers.''}

To overcome the limitations of manual key presses, some VTubers have used software features that automatically trigger certain expressions when their facial movements exceed pre-defined thresholds.
For example, some set the software to trigger a smiling expression when the mouth moves upward past a certain point.
P15 commented, \textit{``Automatic is more convenient and natural. It feels like a real facial expression since I can trigger the avatar's expression through facial movements that relate to it, making it more immersive than pressing a button.''}
However, this method has its own challenges.
Setting the threshold too low could trigger superfluous expressions while setting it too high required exaggerated facial movements, potentially causing physical stress.
P1 explained, \textit{``For instance, if I need to pout to change an expression, I have to exaggerate the movement to cross the threshold, making my mouth sore. Furthermore, as an Asian with racially smaller eyes, triggering expressions relying on eye movement often fails, making automatic control less convenient.''}

\subsubsection{Hand Control}
VTubers believe that hand animations enhance the liveliness of their avatars and enrich their content.
P5 and P16 mentioned that having the avatar's hands move while eating or playing an instrument would add a greater sense of realism.
Additionally, P4 highlighted the potential for humor by having the avatar perform gestures, such as raising the middle finger, to entertain viewers.
Despite these positive views on hand control, it is primarily used in full-body setups, with most participants either avoiding or discontinuing hand control in upper-body setups.

VTubers using 2D models pointed out that implementing hand control requires additional rigging for the hands.
As mentioned in Section \ref{subsubsec:designspace-modeltype}, 2D models necessitate manual rigging of each body part, leading to higher expenses as the rigging scope expands.
P2 emphasized that, given these expenses, prioritizing natural facial rigging is more beneficial, relegating hand rigging to a lower priority.
On the other hand, VTubers using 3D models with rigged hands noted that setting up the necessary software for hand control in upper-body setups poses a significant challenge.
As mentioned in section \ref{subsubsec:avatar-setup-upperbody}, the fragmented nature of VTubing equipment requires VTubers to operate multiple programs and navigate complicated configurations, complicating hand control.
P5 illustrated this by saying, \textit{``Each program has functions a, b, and c, but unfortunately, no software combines all these features. Therefore, I have to prepare the stream by launching multiple programs and orchestrating them while avoiding crashes, which is both cumbersome and complex.''}

Challenges with hand control also stem from hardware limitations.
When using RGB or RGB-D cameras, VTubers must ensure that their hands remain within the camera’s field of view and do not interfere with face tracking.
P9 explained, \textit{``If my hands even slightly cover my face, the avatar's expression becomes unnatural, so I constantly have to pay attention, which is inconvenient. I turn off hand tracking when I am not using it.''}
P1, who used an RGB-D camera, added, \textit{``The range of motion is limited. If I extend my arms too far, they are not recognized, so managing this throughout the stream is bothersome.''}
Despite these limitations, few participants opted to use higher-fidelity devices like data gloves.
P8 mentioned, \textit{``Data gloves are prohibitively expensive and, even when used, do not deliver the desired performance. In addition, current off-the-shelf VTubing software does not support data gloves, making the existing setup the most practical option.''}

\subsubsection{Body Control}
To ensure that their movements are accurately reflected in their avatars, VTubers often exaggerate their actions, similar to their approach to facial expressions.
P8 noted, \textit{``Subtle movements make the avatar's joint appear stiff because the motion capture device does not capture them well. You have to exaggerate your movements to make your avatar look realistic.''}
P9 explained that performing as a VTuber involves directing the avatar with exaggerated movements rather than merely reflecting natural ones.

Exaggerated acting with the equipment posed significant physical challenges for VTubers in streams.
P8 compared it to carrying 3-4 kg on the body, describing it as hard labor that caused sweating and shortness of breath after hours of streaming.
This strain made long sessions with a full-body setup difficult.
For example, P9 mentioned that she could stream for 4-6 hours with an upper-body setup but struggled to last more than an hour with a full-body setup.
VTubers also noted that loosening gear for comfort led to incorrect avatar movements, making it necessary to wear the equipment tightly.
Despite enduring these inconveniences, VTubers pointed out that motion trackers cannot properly express rapid movements such as rolling a foot, thus limiting their body control.
P3 noted, \textit{``Moving too quickly can sometimes cause the avatar to adopt unnatural, octopus-like postures. Such errors have occurred a few times during live streams. While my viewers found it amusing, I felt unsettling and embarrassing. That's why I usually avoid full-body tracking during streams. I think the avatar looks the most polished when only the upper body is shown.''}
\subsection{Avatar Interaction}
VTubers interacts with their viewers, other streamers, and the physical world.
This section explores how VTubers use their avatars to interact with different entities, the limitations they encounter, and their needs.

\subsubsection{Interaction with Viewers}
\begin{figure*}
    \centering
    \includegraphics[width=.49\linewidth]{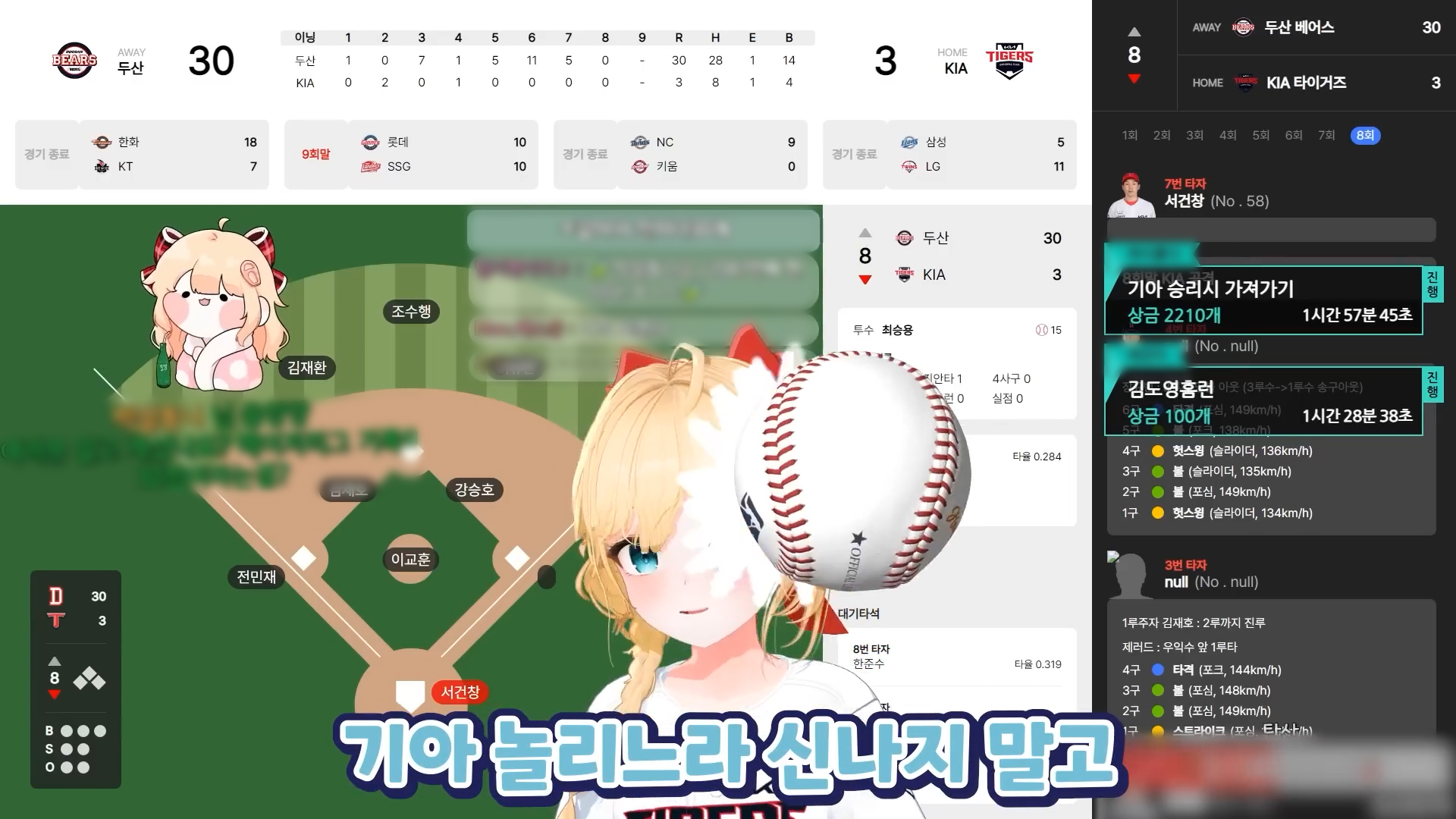}
    \includegraphics[width=.49\linewidth]{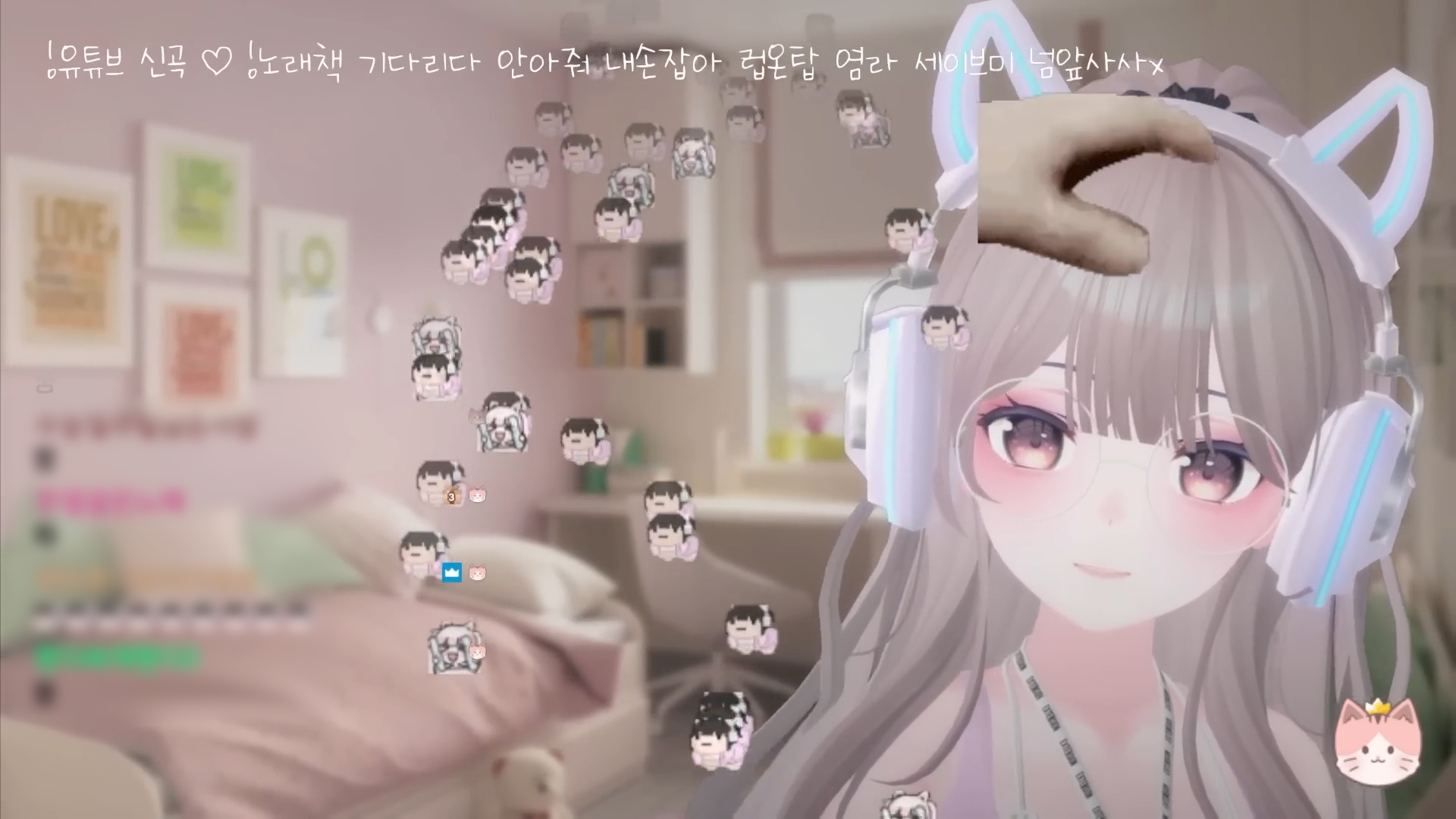}
    \caption{Examples of viewer interaction (left: throwing an object to VTuber \cite{Interaction-Throwing}, right: displaying emojis and stroking VTuber's head \cite{Interaction-Emoji})}
    \Description{This figure shows two examples of interactive features that allow viewers to engage with VTubers during live streams. Left Side (Throwing an Object to VTuber): This image shows a baseball broadcast interface with a VTuber character overlaid in the bottom left corner. The VTuber is a cartoonish character with blonde hair and cat ears. A large baseball is being thrown toward the character, simulating the interaction where viewers can "throw" virtual objects at the VTuber during the stream. The on-screen text, "Don’t get too excited about teasing Kia Tigers (professional baseball team in Korea)." This interaction engages the audience playfully and dynamically, making it feel like they can affect the streamer's experience. Right Side (Displaying Emojis and Stroking VTuber’s Head): This image illustrates another interaction where viewers can display emojis on the screen and virtually pet the VTuber’s head. The VTuber on this side is a 3D model with long gray hair, cat ears, and a headset. Several identical mini emojis of the VTuber are floating around the screen. Additionally, a virtual hand is shown petting the VTuber’s head, creating a personal and engaging interaction that mimics real-world actions.}
    \label{fig:viewer-interaction}
\end{figure*}

Similar to real-person streamers, VTubers interact with their viewers through various features, such as chat, donations, polls, and roulettes.
Additionally, VTubers use their avatars as a unique medium for interaction, creating distinct engagement experiences.
One popular method involves integrating software like the Twitch Integration Throwing System \cite{TwitchThrowing} into their streams, which allows viewers to throw virtual objects at the avatar through donations (Figure \ref{fig:viewer-interaction} left).
When an object hits the avatar, the VTuber reacts exaggeratedly, enhancing the viewers' immersion.
P3 explained, \textit{``When I say something silly, viewers throw things, creating funny situations.''}
VTubers also allow for diverse interactions---changing the avatar’s clothes, dropping emojis on the stream screen, or stroking the avatar’s head---to encourage more interactive viewer engagement (Figure \ref{fig:viewer-interaction} right).
P9 noted, \textit{``When emojis are displayed around the avatar, viewers find it entertaining and get more involved. Even those who were not chatting before trying it out,''} highlighting the positive impact of interactive experience.

Even after their streams, VTubers actively engaged with their viewers.
Like real-person streamers, they used social media platforms such as YouTube, Discord, and X to attract new fans and strengthen their communities.
P14 emphasized the importance of maintaining connections with viewers even when not streaming, stating, \textit{``It is crucial to keep in touch with the audience so they look forward to the next stream and have something to anticipate even when the stream is offline.''}
Interestingly, despite the heavy use of avatars during streams, many VTubers preferred text-based communication after streaming.
They found creating post-stream content with avatars burdensome, viewing it as an extension of their streaming work. 
For instance, taking a selfie of an avatar required turning on the computer, launching the software, setting up the tracking equipment, and going through rendering and compositing processes.
P2 commented, \textit{``It is too cumbersome to create a single image by animating the avatar with multiple devices and software and capturing it. It is easier just to start another stream.''}
Nevertheless, participants acknowledged the positive impact of avatar-based interaction and recognized that fans enjoyed seeing VTubers outside streams.
P9 noted, \textit{``Posts that include virtual avatar photos or captures of stream get much better responses than just text on the YouTube community.''}
Similarly, P8 mentioned, \textit{``Viewers often say that three hours of streaming is not enough, and they want to see more of the avatar in other interaction spaces.''}

\subsubsection{Interaction with Other Streamers}
\begin{figure*}
    \centering
    \includegraphics[width=.49\linewidth]{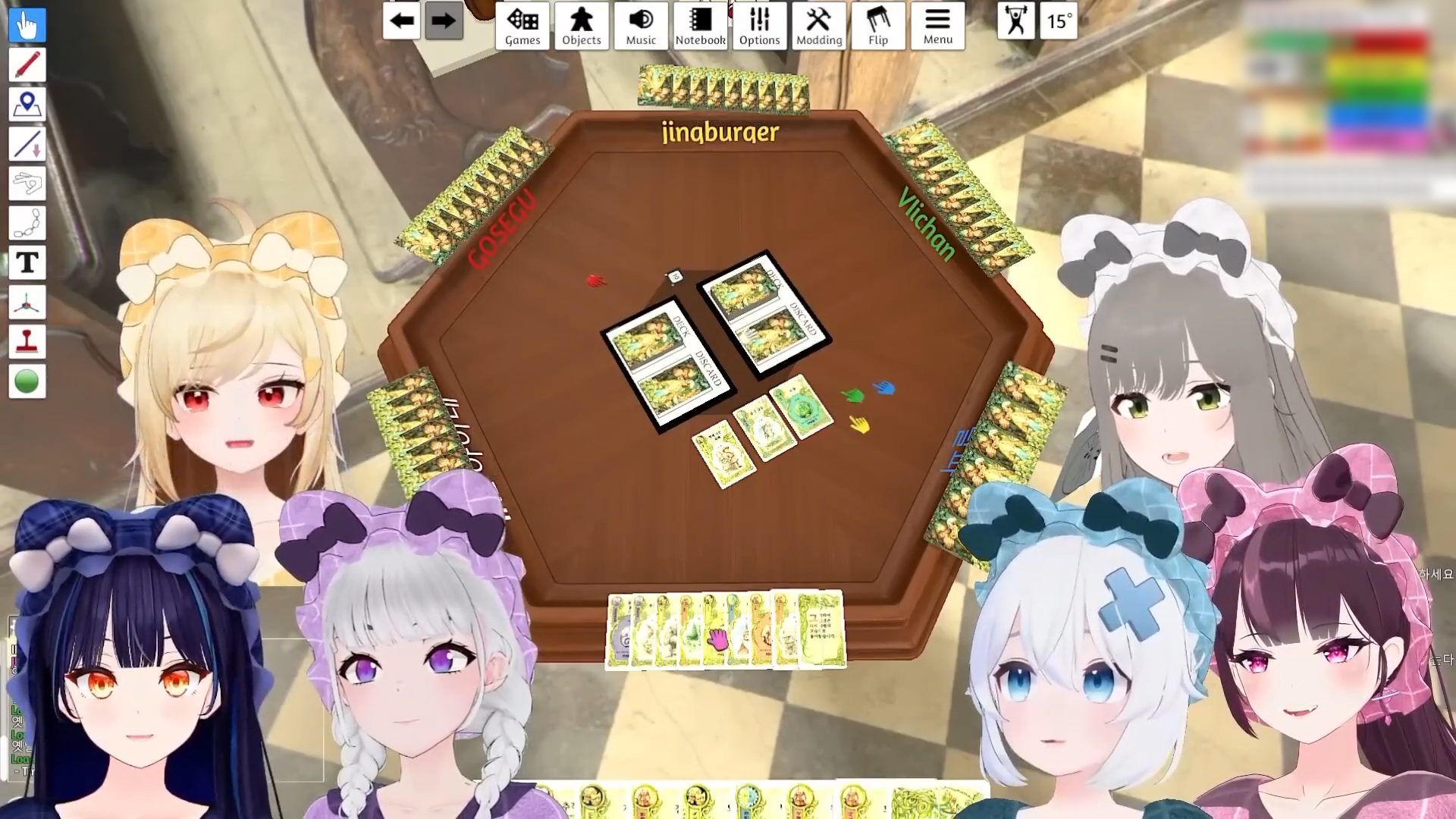}
    \includegraphics[width=.49\linewidth]{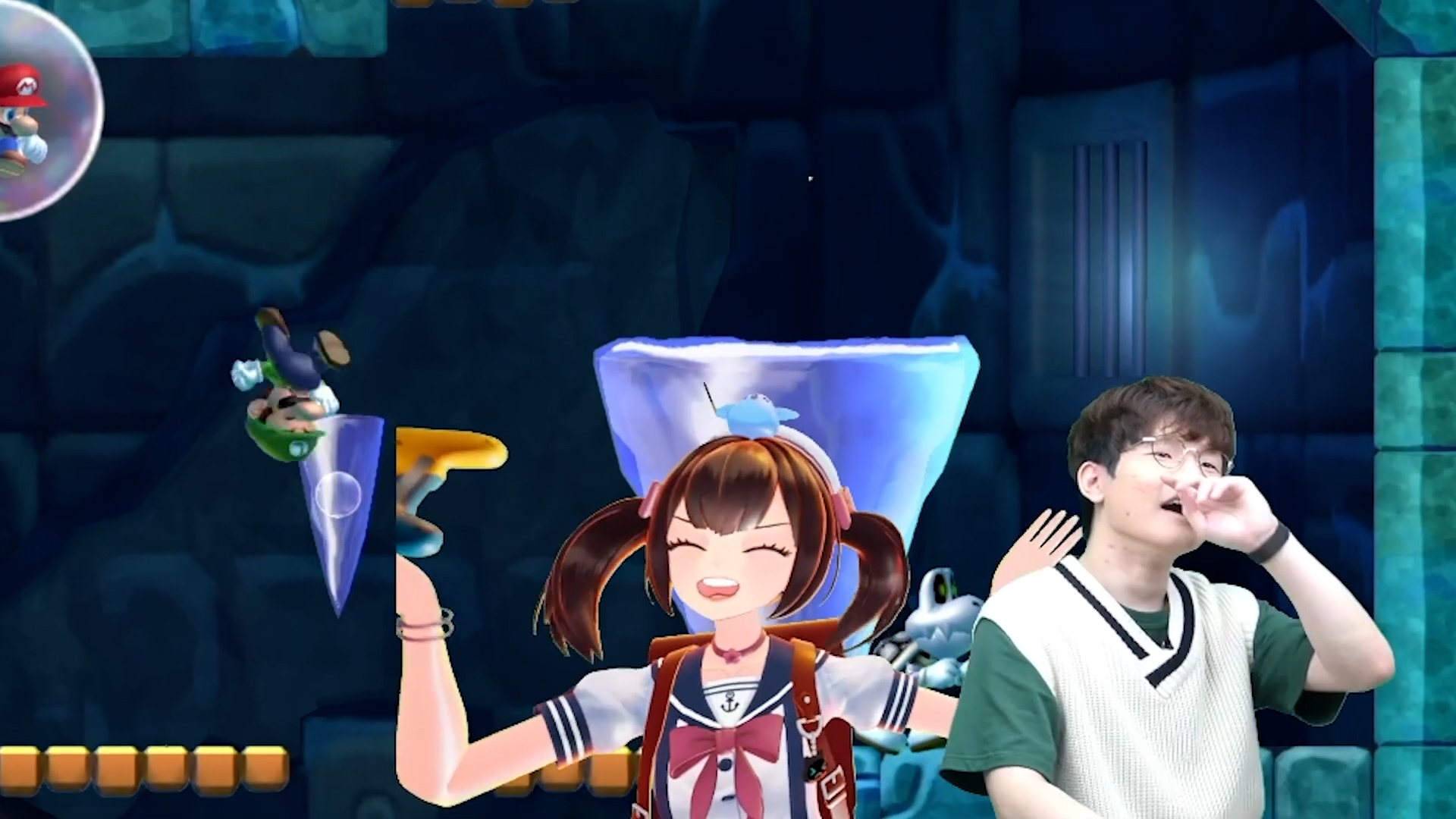}
    \caption{Collaborative streaming with other VTubers and real-person streamers (Image retrieved from \cite{Collaboration-VTuber, Collaboration-nonVTuber})}
    \Description{This figure illustrates examples of collaborative streaming between VTubers and real-person streamers. Left Side (Collaborative Streaming with Other VTubers): This image depicts a virtual tabletop game with multiple VTuber avatars around the board. The avatars represent anime-style characters with distinct appearances, including a blonde-haired VTuber, one with purple twin tails, and others with varying hair colors and styles. Each VTuber's avatar is positioned around the digital game board, indicating their participation in a collaborative, multiplayer game. The tabletop game interface is displayed in the center, with cards and other game elements visible. The VTubers engage in the game together, creating a shared virtual environment for collaboration. Right Side (Collaborative Streaming with Real-Person Streamers): This image shows a collaborative stream featuring a VTuber and a real-person streamer. Next to the VTuber, the real-person streamer appears on screen, smiling and playfully covering their face while participating in the same stream.}
    \label{fig:collaboration}
\end{figure*}

VTubers often engage in collaborative streaming with other VTubers.
Instead of meeting in person, they connect virtually using social VR platforms like VRChat \cite{VRChat} or programs like Discord and OBS Ninja \cite{OBSNinja} to share their avatars’ images in non-VR settings. 
These tools allow multiple VTuber avatars to be displayed in a single stream, as illustrated in Figure \ref{fig:collaboration} left. 
To ensure natural interactions, VTubers prearrange the positioning of their avatars and rehearse their interactions before going live, creating the illusion that they share the same physical space.
P8 noted, \textit{``When preparing for collaborative streams with unfamiliar streamers, I spend several hours a day for a week discussing plans to create more naturally-looking interactions. Furthermore, I hold rehearsal sessions to practice interactions and thoroughly check technical configurations.''}
Despite these preparation, VTubers often feel awkward interacting with each other while staring into a space where no one physically exists, and communication is challenging due to the limited nonverbal expressions of the virtual avatars.
For instance, P3 mentioned that avatars' limited facial expressions make it difficult to time conversations and accurately gauge others’ emotions.
Furthermore, simulating physical interactions, such as avatars making contact, is complicated by the absence of tactile feedback, adding complexity to the collaboration.

VTubers also collaborate with real-person streamers (Figure \ref{fig:collaboration} right).
In these instances, they often prefer to display only an avatar illustration on the screen and interact primarily through voice rather than fully integrating the avatar into the real-world setting.
This preference stems from the difficulty of making an avatar appear natural alongside a real person.
P1 expressed challenges in compositing the avatar naturally into the real-person streamer’s stream. 
Similarly, P16 explained that the differences in background and lighting could make the 3D model look out of place, possibly leading to the sloppy visual quality of streams.

\subsubsection{Interaction with the Physical World}
\begin{figure}
    \centering
    \includegraphics[width=\linewidth]{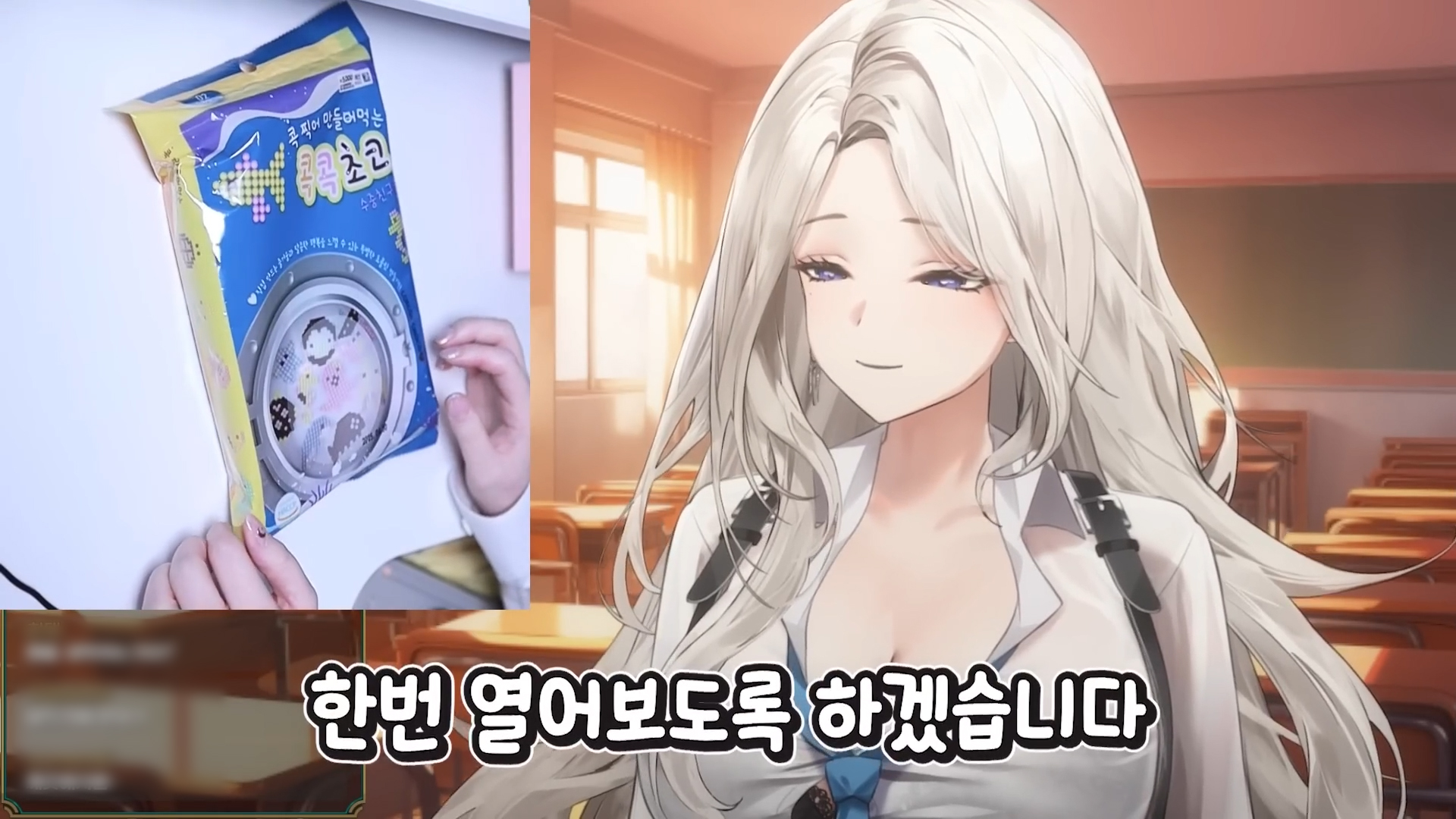}
    \caption{Example of the handcam technique (Image retrieved from \cite{Interaction-Handcam})}
    \Description{This figure demonstrates the "handcam" technique, where a VTuber incorporates real-time footage of their hands interacting with physical objects during a live stream alongside their virtual avatar. Left Side (Handcam View): The left side of the figure shows the handcam footage, which features a pair of hands holding a colorful package. The hands are about to open the package containing small toys for crafting. Right Side (VTuber Avatar): The right side displays the VTuber's avatar, a 3D character with long, blonde hair and a soft, smiling expression. The avatar is presented in front of a classroom background.}
    \label{fig:interaction-handcam}
\end{figure}

VTubers can interact with virtual objects, but interactions with real-world objects, such as eating food, are rare.
P16 noted, \textit{``Even if I try a new hamburger or drink a Starbucks beverage, the viewers cannot tell,''} highlighting the disconnect between virtual avatars and real-world actions.
P15 further explained that this limitation makes it difficult for VTubers to create content related to real life, such as fashion, travel, and food, unlike real-person streamers.

To bridge this gap, some VTubers use a ``handcam'' technique to show their real hands interacting with physical objects alongside their avatar (Figure \ref{fig:interaction-handcam}).
However, this method raises privacy concerns, e.g., \textit{``If my face accidentally reflects off a container while cooking, it could be a serious issue.''} (P8).
Similarly, P11 avoided using a handcam due to the potential risks of revealing personal details, such as tattoos.
Interestingly, some fandoms forbid VTubers from revealing any part of themselves.
P7 observed that some fans strongly oppose even a glimpse of the real person, aligning with findings from Lu et al. \cite{Lu-2021-MoreKawaii}, saying, \textit{``If you show yourself behind the avatar, you are not a VTuber anymore.''}
\section{Discussion}
\label{sec:discussion}
Through a comprehensive analysis of online resources and interviews with professional VTubers, we elaborated on the current state of VTubing equipment and described the current practices and experiences of VTubers, ranging from avatar creation to operation and interaction. 
This section highlights the unique challenges and insights with respect to VTubing and relates our findings to the previous work.

\subsection{Avatar Design \& Creation}
Research on virtual influencers suggests that perceived anthropomorphism and consistent textual and visual cues play a significant role in building audience trust and engagement \cite{Gerlich-2023-PowerOfVI, Yoo-HowCanI-2024}. 
In the case of VTubers, aligning an avatar's appearance, voice, and personality can be a powerful way to enhance authenticity and reduce the psychological distance between performers and their audience.
The following sections explore the unique challenges involved in constructing avatars.

\subsubsection{Appealing vs. Sustainable Identity}
VTubers enjoy significant creative freedom when designing their avatars and identities, often with few limitations.
As Xu and Niu noted \cite{Xu-2023-Understanding}, VTubers are encouraged to design their identities with a target audience in mind, tailoring their avatars to suit different viewer preferences.
However, our findings reveal that this freedom can pose challenges.
Similar to the experience of designers \cite{Koch-2020-Semanticcollage}, many VTubers reported feeling a creative burden when developing an avatar from scratch.
Some VTubers derive their avatar concepts from their authentic selves, while others begin streaming without fully developing their virtual identities.
This often results in an organic rather than a strategically crafted persona.

While creating an identity that resonates with a broader audience is crucial, deviating too far from one’s authentic self can lead to challenges in sustaining the persona over time.
Although the gap between a VTuber's initial concept and actual identity may gradually narrow over time \cite{Lu-2021-MoreKawaii}, our findings suggest that an wide gap can negatively impact a VTuber's well-being.
The mental burden of maintaining a vastly different character can lead to burnout, negatively impacting long-term streaming efforts.
This aligns with research indicating that avatar users in virtual worlds are more satisfied when the psychological gap between themselves and their avatars is smaller \cite{Ducheneaut-2009-Body}.
Therefore, VTubers must balance creating an appealing, engaging persona and ensuring it can be sustained for the duration of their VTubing career.

\subsubsection{Harmonizing Avatar Appearance and Voice}
The harmony between an avatar’s appearance and voice influences the believability of the character \cite{Lam-2023-Effects}, with mismatches potentially leading to the uncanny valley effect \cite{Higgins-2022-Sympathy}.
Xu and Niu observed that viewers of VTubers place greater emphasis on voice compared to real-person streamers, recognizing it as a critical factor in overall appeal \cite{Xu-2023-Understanding}.
In our study, VTubers were mindful of this relationship, either designing avatars that harmonized with their voices or adjusting their vocal performance to align more closely with their avatar’s visual characteristics.
However, there is still a limited understanding of how specific aspects of an avatar’s appearance and voice impact the viewing experience.
While existing research explores user perceptions of the visual appearance and voice of virtual agent avatars \cite{Lam-2023-Effects, Higgins-2022-Sympathy} and game avatars \cite{Kao-2022-AudioMatters}, it may not fully account for the unique dynamics of viewer experience in VTubing contexts.
For instance, Otaku, a primary fan base for VTubers, might prefer a kawaii voice even when the voice and appearance are not perfectly matched \cite{Seaborn-2023-Kawaii, Seaborn-2023-Can}.
Future research could delve deeper into these dimensions to provide more detailed guidance on the optimal design of VTuber avatars and voices.

Furthermore, an exploration of VTubers’ personal views on their voices, beyond viewer perceptions, could be valuable.
Okano et al. found that individuals dissatisfied with their voices experience more positive emotional reactions when their avatars use morphed voices \cite{Okano-2022-AvatarVoice}.
Extending this research to VTubers could reveal how voice preferences affect their overall experience, suggesting new strategies to enhance VTubers' performance.

\subsubsection{Challenges in Creating Original Avatars}
\label{subsubsec:ChallengesinCreating}
Most VTubers lack the skills to create avatars with the originality they envision.
While template-based tools offer affordable and accessible options for those without professional graphic design expertise, their limited customization capabilities often prevent VTubers from achieving a truly unique look.
This issue mirrors findings from previous research on virtual world users who felt constrained in self-expression due to insufficient avatar customization options \cite{Pace-2009-Socially, Neustaedter-2009-Presenting}.
Ducheneaut et al. noted that users were particularly dissatisfied with limited ``hairstyle'' options, as hairstyles serve not only as a means of identity expression but also as a way to be easily recognized during interactions with others \cite{Ducheneaut-2009-Body}.
For VTubers, however, originality goes beyond self-expression---it is essential for differentiating themselves and attracting viewers, making it a key element of their business strategy.
As a result, many VTubers prefer to commission professional artists despite the financial burden this entails.

Kitbashing, or assembling avatars from ready-made assets, provides a more cost-effective alternative to fully custom-made avatars.
This is similar to the tendency of VR/AR creators with limited technical expertise to rely on pre-made assets \cite{Ashtari-2020-Creating, Krauss-2021-Current}.
They find it easier and more reliable to utilize high-quality assets than to build everything from scratch.
However, as discussed in Section \ref{subsubsec:avatar-creation-kitbashing}, kitbashing presents its own challenges, such as ensuring that assets align with desired look and are compatible with one another to avoid avatar malfunction.
Furthermore, kitbashing often requires the use of professional graphics tools which can lead VTubers to seek assistance from skilled artists.
These limitations highlight the need for future research into tools and methods that empower VTubers to create unique avatars independently, fostering greater creative expression without heavy reliance on professional assistance.
\subsection{Avatar Setup \& Control}
VTubers utilize various software and hardware to control their avatars, a process that closely parallels the workflows of VR and CG creators.
For instance, similar to VR creators, VTubers employ HMD devices and diverse authoring tools to produce their content \cite{Ashtari-2020-Creating, Krauss-2021-Current, Nebeling-2018-Trouble}.
Additionally, like motion-capture actors, they manipulate and perform through their avatars \cite{Cheon-2024-Cretive}.
However, the live streaming environment and often limited technical expertise introduce distinctive challenges and characteristics specific to VTubers.

\subsubsection{Challenges in Integrating Equipment}
Integrating equipment presents significant technical challenges for VTubers, exceeding those encountered by real-person streamers. 
While traditional streamers typically combine broadcasting, video, and audio equipment into a single system \cite{Drosos-2022-DesignSpace}, VTubers must also integrate and manage avatar control systems, adding an extra layer of complexity to their setup. 
In addition, the available software and hardware range from consumer-grade to professional-level solutions, creating fragmented environments. 
This fragmentation mirrors the challenges faced by VR/AR creators \cite{Ashtari-2020-Creating, Krauss-2021-Current, Nebeling-2018-Trouble} and imposes a significant cognitive load.

Cognitive load theory (CLT) was initially proposed to optimize learning outcomes by managing the cognitive load imposed on a learner’s limited working memory \cite{Sweller-2011-Cognitive}. 
CLT comprises three distinct components---intrinsic, extraneous, and germane cognitive load. 
Previous studies in HCI have sought to integrate usability principles with CLT, underscoring the importance of reducing the complexity of extraneous environments to minimize unnecessary extraneous cognitive load \cite{Hollender-2010-Integrating}. 
In the context of VTubing, simultaneously managing and integrating fragmented devices can introduce excessive extraneous cognitive load and decrease overall usability, ultimately hindering performance. 
We further discuss these cognitive issues in subsequent sections and propose several future system designs in Section \ref{sec:design opportunities} to alleviate these burdens and enhance user performance.

The lack of flexible support for mixing hardware across different software further complicates the setup process, often necessitating additional software to ensure system compatibility.
These issues are particularly pronounced in mixed VR setups, where equipment from different manufacturers is frequently incompatible.
As a result, the technical skills required for VTubers have become a more critical factor \cite{Pellicone-2017-Performing} compared to real-person streamers.
VTubers lacking these skills face limitations in creating diverse content and fully expressing themselves through avatars.
These technical barriers point to a need for future research that identifies the factors contributing to technical demands during avatar setup and explores solutions to simplify these processes, allowing VTubers to focus more on content creation.

\subsubsection{Technical Limitations in Hand and Body Control}
VTubers encounter notable technical limitations when controlling hand and body movements compared to facial control.
Many VTubers rely on RGB/RGB-D cameras, HMDs with controllers, and motion trackers to manipulate their avatars' hand and body movements.
While these devices are more affordable than data gloves or motion capture suits, they impose significant constraints during streams.
For instance, RGB/RGB-D cameras require VTubers to remain within a limited tracking range, and motion trackers must be fastened tightly for optimal tracking quality.
Furthermore, VTubers should exaggerate their movements to achieve expressive and accurate portrayals, posing additional physical challenges.
These challenges are comparable to those employed by professional motion capture actors in the animation and film industries \cite{Cheon-2024-Cretive}.
However, VTubers typically use motion capture devices with lower fidelity than those employed in professional production environments, which may introduce further performance challenges.
For instance, the nature of HMDs, including excessive weight \cite{Li-2019-Measuring}, detachment from the physical environment \cite{sarupuri-2024-dancing}, and motion sickness \cite{Wu-2023-Interactions}, can exacerbate these physical difficulties.
Given these challenges, future research could explore cost-effective improvements in hand and body control, providing VTubers with greater freedom of movement while reducing physical strain.
In addition, investigating the labor intensity of VTubers' performances in relation to devices across a spectrum of fidelity levels would provide valuable insights into improving their experiences.

\subsubsection{Cognitive Load during VTubing Performances}
VTubers, like real-person streamers, aim to entertain their audience through performance \cite{Hamilton-2014-StreamingOnTwitch}.
To compensate for the limited expressiveness of their avatars, VTubers often rely on exaggerated voices \cite{Wan-2024-Investigating}, facial expressions, and gestures---commonly referred to as overacting---to enhance the viewing experience.
They also strive to selectively playback predefined animations with dramatic expression, such as shedding tears or displaying sparkling, heart-shaped eyes, that motion capture technology cannot achieve.
VTubers recognize that, while they cannot appear as realistic as real-person streamers, the ability to incorporate unrealistic and animated expressions is a unique advantage that sets them apart.
However, manually triggering these animations can significantly increase cognitive load, particularly when performing content that requires high levels of concentration. 
In some cases, VTubers even opt to forgo avatar control to focus entirely on their content.
This highlights how the synchronicity and ephemerality of live streaming uniquely impact VTuber performances.
While real-person streamers and moderators also face challenges such as excessive multitasking and information overload \cite{Cai-2021-Moderation, Hamilton-2014-StreamingOnTwitch, Wohn-2020-AudienceManagement}, they are rarely forced to make the extreme trade-off between abandoning their performance and focusing solely on content delivery.
Some VTubing software includes automation features similar to those proposed by Tang et al. \cite{Tang-2021-AlterEcho}, which may help alleviate this burden.
However, our findings suggest that these systems can introduce other issues related to setting thresholds for automatic triggering, as described in Section \ref{subsec:facial-control}. 
We will discuss related design opportunities in Section \ref{sec:design opportunities}.

\subsection{Avatar Interaction}
Compared to real-person streamers, VTubers are more likely to traverse various media platforms at the intersection of virtual and real worlds \cite{ferreira-2022-concept, Regis-2024-VTubers}. 
Through avatars, they form unique interactions with viewers, streamers, and the physical world.
However, limitations in current technology and the complexity of content production often constrain these interactions.

\subsubsection{Differences in Avatar Usage during and after Streaming}
Real-person streamers build parasocial relationships with their audiences through attributes like trustworthiness, attractiveness, and expertise, significantly impacting viewer loyalty and engagement \cite{Kim-2023-Parasocial, Hou-2024-Understanding}.
VTubers, operating through virtual avatars, face unique challenges in establishing similar levels of authenticity and trust. 
However, the alignment between avatar appearance and user behavior, along with strategic use of interactive tools like audience-triggered animations, can bridge this gap and deepen viewer engagement. 
Future systems should integrate features that enhance the expressiveness and adaptability of avatars to foster stronger parasocial connections.

Outside of streams, however, VTubers often find it burdensome to use their avatars for social media interactions.
Unlike real-person streamers, who can easily share selfies, VTubers must navigate the complex process of combining multiple hardware and software systems to control and render their avatars.
This technical complexity discourages VTubers from creating avatar-based content after streams.
This practice contrasts with virtual influencers, who primarily produce high-quality, non-real-time content for platforms like Instagram \cite{Choudhry-2022-Felt}.
Virtual influencers are typically created and managed by professional teams equipped with advanced CGI technologies \cite{Conti-2022-VIs}, underscoring how access to technical resources influences the choice of content formats and platforms.
Future research into simplifying avatar content creation processes could maximize avatars' potential to extend engagement beyond VTuber's streaming sessions.

\subsubsection{Reduction in Nonverbal Cues in Collaborations with Other Streamers}
Collaborative streaming, which generates synergy and increases audience engagement \cite{Jia-2020-HowtoAttract}, is beneficial for both VTubers and real-person streamers. 
VTubers can experiment with creative scenarios, such as avatar swaps between VTubers or virtual avatars interacting with real people---possibilities that are not feasible in the physical world \cite{Lu-2021-MoreKawaii}.
However, the use of avatars reduces nonverbal cues, making collaboration more challenging.
This aligns with research on social VR and telepresence, which highlights the difficulties caused by the scarcity of facial cues \cite{Moustafa-2018-longitudinal} and lack of tactile feedback \cite{Jung-2021-Use}.
Previous studies have found that social VR users adopt unique interaction methods, such as using `mirrors,' to compensate for diminished social cues \cite{Fu-2023-MirrorDweller}. 
A significant distinction between VTubers and social VR users, however, is the presence of an audience---`viewers'---in the broadcast context.
VTubers aim to create the illusion that their avatars and those of other streamers coexist and naturally interact within the same virtual space, enhancing immersion in their role-playing performances.
To achieve this, VTubers often rely heavily on performance tricks to compensate reduced information, which increases their cognitive load during collaborative streams.
Future research could focus on enhancing nonverbal expression in avatar-mediated communication, helping to facilitate more natural interactions during these collaborations.

\subsubsection{Difficulties in Interacting with the Real World}
Technological limitations currently prevent seamless interaction between the virtual and real worlds.
For example, avatars cannot naturally reflect interactions with physical objects, creating a dissonance in behavior.
Some VTubers attempt to address this with ``handcam'' strategies, overlaying video of parts of their real body, such as their hands, on top of the streams.
However, this approach carries risks, including unintended privacy leakage caused by reflections.
In addition, some viewers prefer the illusion of a fully virtual persona and may react negatively to any part of the VTuber's real appearance \cite{Lu-2021-MoreKawaii}.
These technical and cultural limitations may place VTubers at a disadvantage compared to real-person streamers when creating branded content to promote products or services.
For example, promoting a physical product as a VTuber often requires the creation of high-quality digital assets representing the product, significantly increasing the cost and time required for content production \cite{Jhang-2024-Analysis}.
Future design research could focus on developing systems that translate real-world interactions into virtual spaces or blend virtual avatars more seamlessly with the real world, enabling more natural and fluid interactions.
\section{Design Opportunities}
\label{sec:design opportunities}

\begin{figure*}
    \centering
    \includegraphics[width=\linewidth] {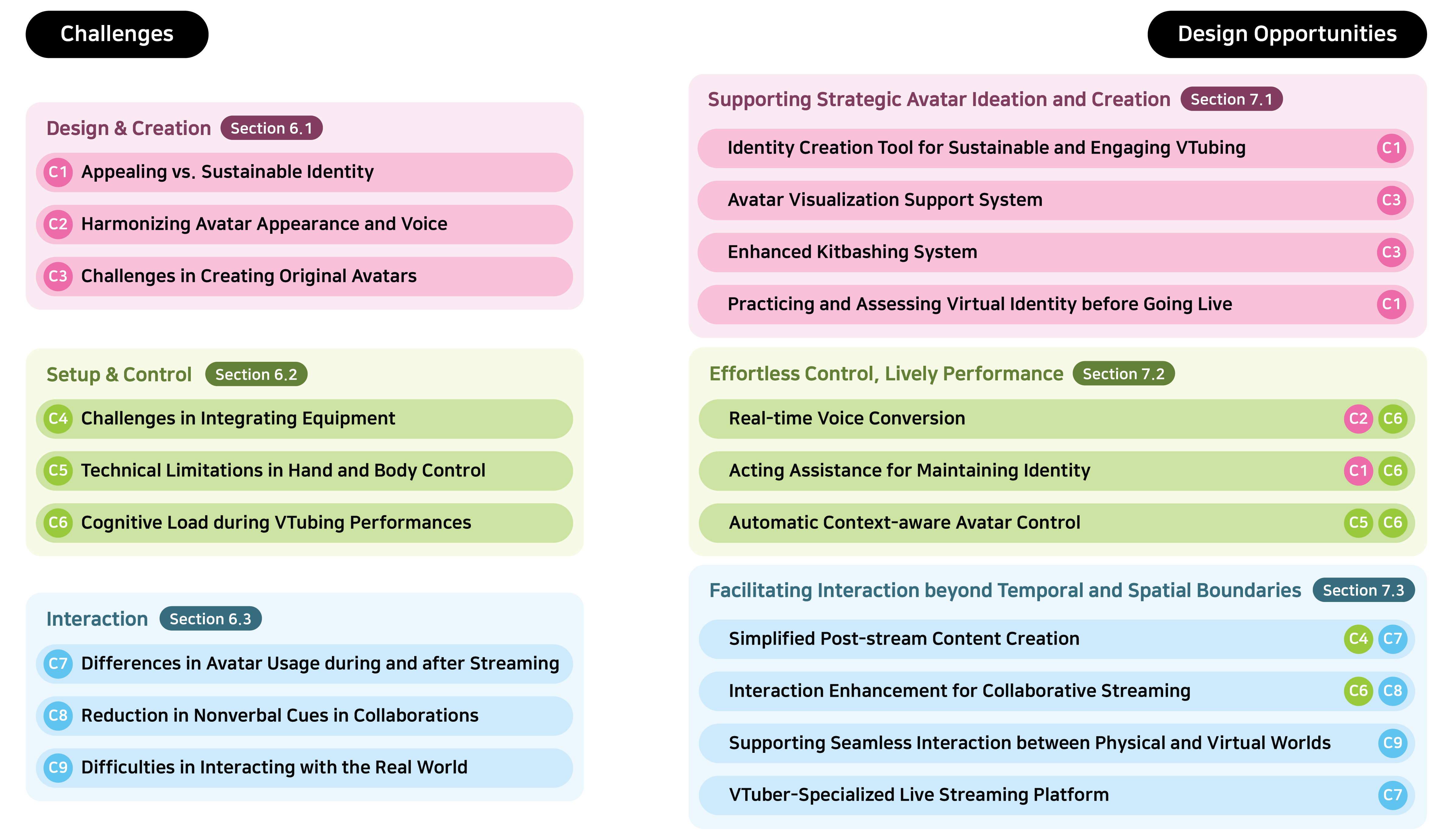}
    \caption{Mapping the relationship between challenges and design opportunities}
    \Description{The image visually maps the relationship between challenges and design opportunities in VTubing. It is divided into three categories of challenges (on the left) and their corresponding design opportunities (on the right). Each category is represented by links through shared identifiers (e.g., C1, C2).
    Challenges (left) include the following categories:
    Design & Creation (Section 6.1): C1: Appealing vs. Sustainable Identity, C2: Harmonizing Avatar Appearance and Voice, C3: Challenges in Creating Original Avatars; Setup & Control (Section 6.2): C4: Challenges in Integrating Equipment, C5: Technical Limitations in Hand and Body Control, C6: Cognitive Load during VTubing Performances; Interaction (Section 6.3): C7: Differences in Avatar Usage during and after Streaming, C8: Reduction in Nonverbal Cues in Collaborations, C9: Difficulties in Interacting with the Real World
    Design Opportunities (right) include the following categories:
    Supporting Strategic Avatar Ideation and Creation (Section 7.1): Identity Creation Tool for Sustainable and Engaging VTubing (C1), Avatar Visualization Support System (C3), Enhanced Kitbashing System (C3), Practicing and Assessing Virtual Identity before Going Live (C1); Effortless Control, Lively Performance (Section 7.2): Real-time Voice Conversion (C2, C6), Acting Assistance for Maintaining Identity (C1, C6), Automatic Context-aware Avatar Control (C5, C6); Facilitating Interaction Beyond Temporal and Spatial Boundaries (Section 7.3): Simplified Post-stream Content Creation (C4, C7), Interaction Enhancement for Collaborative Streaming (C6, C8), Supporting Seamless Interaction between Physical and Virtual Worlds (C9), VTuber-Specialized Live Streaming Platform (C7).
    Shared identifiers visually link challenges to opportunities, highlighting how specific challenges influence the proposed design solutions. This mapping emphasizes the thematic connections between the two aspects and their relevance in addressing VTuber challenges.}
    \label{fig:Discussion-DesignOpportunity}
\end{figure*}

Based on the challenges identified, we present various design opportunities to enhance the VTubing experience.
These opportunities are categorized into three key areas: (1) supporting strategic ideation and creation, (2) enabling effortless control and lively performance, and (3) facilitating interaction beyond temporal and spatial boundaries.
Figure \ref{fig:Discussion-DesignOpportunity} illustrates the relationship between the discussed challenges and the proposed design opportunities.

\subsection{Supporting Strategic Avatar Ideation and Creation}

\subsubsection{Identity Creation Tool for Sustainable and Engaging VTubing}
VTubers can design their virtual identities with potential target audiences in mind, but focusing solely on audience preferences may make it difficult to sustain these personas over time.
A system that helps VTubers create identities that are both appealing to viewers and sustainable for long-term use would benefit novice VTubers by enabling strategic performance design and supporting growth.
For example, such a system could help VTubers to define target audience segments, analyze viewer preferences, and generate relevant keywords for persona development \cite{Nandapla-2020-Customers}.
It could also suggest ways to incorporate the VTuber’s authentic identity with traits that resonate with audiences.
With these recommendations, VTubers could design diverse virtual identities that balance personal fulfillment with audience engagement.

\subsubsection{Avatar Visualization Support System}
When creating avatars, VTubers often collect visual references to conceptualize their appearance and communicate their ideas effectively to professional artists.
A generative AI-based tool like GANCollage \cite{Wan-2023-GANCollge} could support this visualization process in a more creative way.
For example, a system could recommend suitable visual features---such as art style, color schemes, facial features, hairstyles, and accessories---based on the VTuber's virtual identity and chosen visual references \cite{Wang-2023-PopBlends}.
VTubers could then select preferred features from these recommendations, generate visuals, and then combine them to build their desired appearance.
This process would help VTubers develop a clearer vision of their avatar while improving collaboration with stakeholders, ultimately ensuring that the avatar aligns with their preferences.

\subsubsection{Enhanced Kitbashing System}
A system designed to improve the kitbashing process could address the challenges mentioned in Section \ref{subsubsec:ChallengesinCreating}, making it easier and faster for VTubers to create high-quality 3D avatars without advanced technical skills.
For instance, the system could suggest compatible 3D assets from marketplaces based on descriptions or sketches about an avatar \cite{Liu-2019-Learning}.
If certain elements do not perfectly align with the VTuber's vision, the system could offer options for fine-tuning or generating new components \cite{Qi-2024-Tailor3d,Raj-2023-Dreambooth3d}.
To solve compatibility issues, the system could also automatically adjust assets to fit the avatar's body, preventing rigging problems and ensuring a natural look \cite{Lund-2023-AI}.
These features would streamline avatar creation, allowing VTubers to focus on originality and creativity while minimizing technical barriers.

\subsubsection{Practicing and Assessing Virtual Identity before Going Live}
VTubers' identities often evolve as they stream \cite{Lu-2021-MoreKawaii}.
However, to foster more strategic and sustainable VTubing, it would be useful to provide an environment where VTubers can practice and assess their identity fit before going live.
With recent advancements in AI, it is possible to simulate streaming environments by generating virtual audience interactions, allowing VTubers to rehearse \cite{Cherner-2023-AIPowered, Liao-2018-VRSpeech}.
After each practice session, the system could provide feedback on their performance, evaluating how well they maintained or enhanced their identity \cite{Glemarec-2022-controlling}.
This support help VTubers refine their avatar’s persona and deliver a more polished, authentic performance to their audience.

\subsection{Effortless Control, Lively Performance}

\subsubsection{Real-time Voice Conversion}
Recent advancements in voice conversion technology have created new opportunities for real-time streaming, offering high-quality voice changer options \cite{Chen-2024-Conan's, Supertone}.
Applying this technology in the VTubing context allows VTubers to design voices that align with their visuals and virtual identities.
For example, one can imagine a system that recommends the characteristics of a voice changer based on the avatar’s appearance, virtual identity, and the target audience’s preferences \cite{Tian-2024-User}.
Additionally, the system could offer features for fine-tuning, such as blending the converted voice with the VTuber's natural voice, adjusting pitch, and modifying intonation \cite{Byeon-2023-AVOCUS}.
These features would allow VTubers create unique voices without the pressure of voice-acting, enabling them to produce more creative content that perfectly complements their avatar.

\subsubsection{Acting Assistance for Maintaining Identity}
To help VTubers perform a variety of identities more naturally, systems that assist with acting could be developed.
These tools could create personalized databases or wikis for each VTuber, organizing vocabulary and background knowledge tailored to their persona.
For instance, a VTuber acting as a medieval vampire persona could use a tool that archives historical terminology and vampire lore, ensuring consistency in their language.
Additionally, the system could analyze the streaming context in real-time, suggesting dialogue that aligns with the persona and providing feedback on how consistently the VTuber stays in character \cite{Yang-2024-Simschat, Jang-2022-CfCC, Ashby-2023-Personalized}.
Designing such tools would foster creativity and support more dynamic and immersive VTubing performances.

\subsubsection{Automatic Context-aware Avatar Control}
VTubers often trigger predefined animations to enhance their avatars' expressiveness, but this manual process can cause significant cognitive and physical strain.
To reduce the load on VTubers, future research could explore systems that automatically trigger appropriate animations by analyzing data such as chat messages, audio cues, and on-screen interactions \cite{Shapiro-2019-Ubebot, Yen-2023-StoryChat, Tang-2021-AlterEcho, Armstrong-2024-SealMates}.
Technologies like text-to-motion \cite{Zhang-2024-MotionDiffuse, Guo-2022-GeneratingMotions, Azadi-2023-MakeAnAnimation} and co-speech motion generation \cite{Liu-2024-CoSpeechMotion, Bhattacharya-2024-Speech2UnifiedExpresion, Wang-2024-MMoFusion} could be leveraged to create systems that generate context-specific animations based on chat inputs or the VTuber’s voice.
Automating avatar control in this way would enhance expressiveness while significantly reducing the effort required from VTubers, allowing for more fluid and engaging performances.

\subsection{Facilitating Interaction beyond Temporal and Spatial Boundaries}

\subsubsection{Simplified Post-stream Content Creation}
VTubers face significant challenges when creating avatar-based content after streaming due to the complex processes involved, such as motion capture and rendering, which can be as demanding as starting another VTubing session.
Generative AI technology has the potential to streamline these tasks and significantly reduce the workload for VTubers.
By fine-tuning generative AI models such as Stable Diffusion \cite{Rombach-2022-LatentDiffusion, StabilityAI-2024-SD3} with the VTuber’s avatar, they could quickly generate images or videos with desired poses or compositions. 
These tools would allow VTubers to easily create content that engages fans across multiple channels, enhancing their branding and extending audience interaction beyond the stream.

\subsubsection{Interaction Enhancement for Collaborative Streaming}
During collaborative streams with other creators, VTubers often face communication challenges due to the reduction of non-verbal cues, which can lead to awkward interactions.
Drawing on social VR research, technologies like bio-signal visualization \cite{Lee-2022-Understanding} and haptic feedback \cite{Fermoselle-2020-Let's} could help address the lack of non-verbal cues in avatar-mediated communication.
Additionally, using physics simulations to prevent unnatural movements, such as avatar penetration during interactions \cite{Sugimori-2021-AvatarTracking, Sugimori-2023-AvatarTracking}, and generating real-time sound effects \cite{Yang-2023-Diffsound} could further reduce the performance strain on VTubers while enhancing viewer engagement and immersion.

\subsubsection{Suppoting Seamless Interaction between Physical and Virtual Worlds}
VTubers face often challenges when trying to interact with physical objects or environments while remaining in their virtual space.
Synchronizing virtual avatars with the physical world or integrating physical objects into virtual environments could open up new possibilities for content creation.
Advanced technologies, such as object segmentation \cite{Kirillov-2023-SegmentAnything,Zhao-2023-FastSegement} and image-to-3D generation \cite{Long-2024-Wonder3D, Liu-2024-One2345++, Li-2023-Instant3D}, could be employed to recognize and convert physical objects into virtual 3D models, enabling seamless interaction between VTubers and their real-world surroundings.
Conversely, augmented reality could enhance seamless interaction by blending avatars into physical spaces, allowing VTubers to engage with the physical world as real-person streamers do \cite{Shapiro-2019-Ubebot, Wang-2020-AvatarMeeting}.
When integrating these techniques, it is crucial to preserve the anonymity of the VTubers by concealing personal information using automatic methods such as privacy-preserving rendering \cite{Zhao-2022-Privacy}.

\subsubsection{VTuber-Specialized Live Streaming Platform}
As mentioned, VTubers have introduced unique forms of interaction on traditional streaming platforms using third-party software. 
A dedicated live-streaming platform for VTubers could take these interactions to the next level, significantly expanding engagement beyond traditional formats.
This platform could offer VTubers new opportunities for creative interaction while increasing viewer immersion and participation.
By supporting various devices like HMDs, viewers could enter the VTuber's virtual space and experience streams from more flexible perspectives \cite{Josyula-2024-Tactile, Lee-2023-Simulcast, Lee-2023-VirtualAvatarConcert}.
Unlike social VR platforms, this platform could provide asymmetrical permission management, allowing VTubers to restrict viewer actions to ensure smooth stream management \cite{Wohn-2020-AudienceManagement}.
It could also enable viewer-participated content, such as collaborative games \cite{Li-2022-Gulliver's}, or post-stream engagement with AI-powered VTubers that simulate real-time interactions in virtual environments \cite{Josyula-2024-Tactile}.

\section{Limitations and Future Work}
This study provides a comprehensive analysis of the equipment used by VTubers and their experiences and challenges in preparing and operating their setups. 
However, several limitations must be acknowledged when interpreting the findings. 
First, most of our participants were independent (indie) VTubers, and we did not include VTubers affiliated with major agencies like Hololive, Nijisanji, or Vshojo. 
Agency-affiliated VTubers typically receive assistance in avatar design, equipment setup, operation, and marketing, which could lead to experiences that differ from those of indie creators. 
Future research should explore the experiences of VTubers from these leading agencies to provide a more balanced perspective. 
Additionally, this study excluded VTubers who primarily use mobile apps like Reality \cite{Shirai-2019-Reality}, which offer an all-in-one platform for avatar creation and equipment setup. 
The unique challenges and opportunities presented by mobile VTubing platforms remain unexplored in this study. 
Future work could examine the experiences of mobile-based VTubers to capture a more complete picture of VTubing practices. 
Although we included participants from diverse cultural backgrounds, most of our sample was concentrated in North America and Asia, particularly Republic of Korea. 
To broaden the understanding of VTubing practices, future research should seek to include VTubers from underrepresented regions such as Europe, South America, and Oceania. 
This would offer valuable insights into how VTubers in different cultural contexts approach equipment use, avatar creation, and ideation, providing a more global perspective on the challenges and experiences faced in VTubing.

\section{Conclusion}
In this study, we conducted a comprehensive analysis of the equipment used by VTubers, focusing on how they prepare, operate, and manage the challenges associated with these tools. 
Through desk research, we identified the specialized equipment involved in VTubing and validated these findings through surveys and interviews with professional VTubers. 
This methodology allowed us to extend the live-streaming design space by introducing six new dimensions specifically related to avatar creation and control. 
Additionally, our interviews provided valuable insights into how VTubers design and create their avatars, set up and manage equipment during streams, and use their avatars to interact with viewers and their virtual environments. 
By examining these experiences and challenges, we identified several design opportunities to improve the VTubing process. 
As the first in-depth study of VTubing equipment, we hope our findings offer valuable insights for both researchers and practitioners. 
We believe this work can contribute to the continued growth and innovation within the VTubing industry.

\begin{acks}
This research was supported by Culture, Sports and Tourism R\&D Program through the Korea Creative Content Agency grant funded by the Ministry of Culture, Sports and Tourism in 2025 (Project Name: Development of AI-based image expansion and service technology for high-resolution (8K/16K) service of performance contents, Project Number: RS-2024-00395886, Contribution Rate: 80\%). This research was also partly supported by the Graduate School of Metaverse Convergence support program (IITP-2025-RS-2024-00430997, Contribution Rate: 10\%) and Innovative Human Resource Development for Local Intellectualization program (IITP-2025-RS-2022-00156360, Contribution Rate: 10\%) through the Institute of Information \& Communications Technology Planning \& Evaluation(IITP) grant funded by the Korea government(MSIT).
\end{acks}

\bibliographystyle{ACM-Reference-Format}
\bibliography{reference}

\end{CJK}
\end{document}